\documentclass[aps,pre,groupedaddress,twocolumn, amsmath,amssymb,showpacs,
eqsecnum,floatfix]{revtex4}
\usepackage{graphicx}
\begin{document}

% Use the \preprint command to place your local institutional report
% number in the upper righthand corner of the title page in preprint mode.
% Multiple \preprint commands are allowed.
% Use the 'preprintnumbers' class option to override journal defaults
% to display numbers if necessary
%\preprint{}

%Title of paper
\title{Hydrodynamics of superfluids confined in blocked rings and wedges}

% repeat the \author .. \affiliation  etc. as needed
% \email, \thanks, \homepage, \altaffiliation all apply to the current
% author. Explanatory text should go in the []'s, actual e-mail
% address or url should go in the {}'s for \email and \homepage.
% Please use the appropriate macro for each each type of information

% \affiliation command applies to all authors since the last
% \affiliation command. The \affiliation command should follow the
% other information
% \affiliation can be followed by \email, \homepage, \thanks as well.
\author{Chandan Dasgupta}
\email{cdgupta@physics.iisc.ernet.in}
%\homepage[]{Your web page}
%\thanks{}
\altaffiliation{Also at Condensed Matter Theory Unit, Jawaharlal Nehru Centre
for Advanced Scientific Research, Bangalore 560064, India}
\affiliation{Centre for Condensed Matter Theory, Department of Physics, 
Indian Institute  of Science, Bangalore 560012, India}
\author{Oriol T. Valls}
\email{otvalls@umn.edu}
\altaffiliation{Also at Minnesota Supercomputer Institute, University of Minnesota,
Minneapolis, Minnesota 55455}
\affiliation{School of Physics and Astronomy,
University of Minnesota, Minneapolis, Minnesota 55455}

\date{\today}

\begin{abstract}
%OTVN editing and define wedge 
Motivated by many recent experimental studies of non-classical
rotational inertia (NCRI) in superfluid and supersolid samples, we  
present  a study of the hydrodynamics of a superfluid confined 
in  the two-dimensional region (equivalent to a long cylinder) between
two concentric arcs of radii $b$ and $a$ ($b<a$) 
subtending an angle $\beta$, with $0 \le \beta \le 2\pi$.
The case $\beta= 2 \pi$ corresponds to a blocked ring.  We discuss
the methodology to compute the
NCRI effects,  and calculate these effects both for small angular
velocities, when no vortices are present, and in the presence of a vortex.
We find that, for a blocked ring, the NCRI effect is small, and that
therefore there will be a large discontinuity in the moment of
inertia associated with blocking or unblocking circular paths. 
%CDN New lines
For blocked wedges ($b=0$) with $\beta > \pi$, we find an unexpected divergence
of the velocity at the origin, which implies the presence of either a region of normal
fluid or a vortex for {\it any} nonzero value of the angular velocity. 
Implications of our results for experiments on ``supersolid'' behavior in
solid $^4{\rm He}$
%CDNEW removed reference to cold atoms and He in porous media
%, superfluidity in ultracold atomic systems and liquid 
%$^4{\rm He}$ in porous media 
are discussed.
A number of mathematical issues are pointed out and resolved. 
\end{abstract}

% insert suggested PACS numbers in braces on next line
\pacs{47.37.+q, 47.32.Ef}
% insert suggested keywords - APS authors don't need to do this
%\keywords{}

%\maketitle must follow title, authors, abstract, \pacs, and \keywords
\maketitle

% body of paper here - Use proper section commands
% References should be done using the \cite, \ref, and \label commands
\section{Introduction}
\label{intro}
Flow without dissipation is the defining feature of superfluidity. 
Because of this  property the moment of inertia of 
a vessel containing a superfluid is different from (smaller than) 
that when the liquid 
is in the normal state. This effect is largest 
in the
absence of vortices, when superfluid flow is  irrotational. 
The difference between the moments of inertia
when the liquid, confined
by  boundary conditions, is in the normal and superfluid states is known
as the ``non-classical rotational inertia'' 
(NCRI). The occurrence of NCRI is often used as an experimental signature 
of superfluidity. Superfluid hydrodynamics and the resulting NCRI have been 
studied extensively~\cite{fetter} in the past for simple geometries, 
such as spherical, cylindrical or rectangular containers rotating about 
a symmetry axis. Because
of several recent developments, some
of which are briefly discussed below, it has become necessary 
to understand the properties of flow of superfluids in enclosures of more 
complicated geometry. These provide the motivation
for our present study.

Recent observations~\cite{chan1,chan1a,chan2,reppy,kojima,reppyn,chan4} 
of NCRI in torsional oscillation experiments on solid $^4{\rm He}$ have been 
interpreted as the occurrence of a ``supersolid'' phase. This interpretation 
of the experimental results is controversial. There is experimental
~\cite{reppy,balibar} and theoretical~\cite{prokofiev1,prokofiev2} 
evidence suggesting 
that the observed NCRI is due to superfluidity along crystalline defects
such as grain boundaries in a 
polycrystalline sample and networks of dislocation lines. 
Since these extended defects 
form complex disordered structures, calculations of the flow properties and 
the rotational inertia of a superfluid confined in irregular-shaped channels 
are necessary for a quantitative assessment of whether this mechanism is the 
correct explanation of the observed results. In this context, it is important 
to examine whether the superfluid component can flow along continuous closed
paths in the sample.
%OTVN changes here (not marked %CDN) seen anyway and found OK 
Since the geometry of the network of defects
would depend on thermodynamic
variables such as temperature and pressure, and
on the cell geometry, the availability 
of such paths would also depend on  these parameters and conditions. Thus, 
an understanding of the dependence of the NCRI on such variables 
requires, for example, 
a calculation of how the NCRI arising from a blocked ring of 
superfluid 
%CD surrounding the axis of rotation 
changes as the blockage is 
removed. To check whether the observed NCRI is due to the occurrence of
extended superfluidity, the NCRI of samples in which the solid
$^4{\rm He}$ is confined in the annular region between two concentric cylinders
has been measured ~\cite{chan1,reppyn} 
in the presence of a barrier in the annulus 
that prevents possible flow of the superfluid along a closed path surrounding 
the rotation axis (the common axis of the cylinders). The NCRI observed
under these conditions is found to be much smaller than that 
for samples in which the artificial block is not present. The calculation 
just mentioned is obviously relevant for a quantitative 
understanding of the results of such 
experiments. Finally, an understanding of experimental 
results~\cite{kojima,reppyn,chan4} 
on the dependence of the NCRI on the frequency of torsional oscillations 
requires a theoretical analysis of vortex formation and 
critical velocity in superfluids 
confined in irregular-shaped channels.

%otvrr3 this paragraph drastically modified
Our study is also partly motivated by %two other reasons.
%One is
the recent explosion of activity in experimental and theoretical
studies of superfluidity and other quantum
phenomena in trapped, ultracold
atomic systems~\cite{bec1,bec2}.
%CDN2 retained a couple of references because we do mention BEC at a few places
%Various signatures of superfluidity, such as persistent flow, NCRI and
%formation of quantized vortices have been observed in both 
%bosonic~\cite{bec1,bec2,becv,bec3} and fermionic~\cite{fermi} systems. While
%the early experiments on such systems were carried out for traps
%with simple geometry, more 
%recent experiments have begun to explore the properties
%of superfluid condensates in traps with more complex 
%structure. Superfluid flow in a toroidal trap has been observed 
%recently~\cite{bec3}, and experimental conditions under which a ring-shaped
%optical trap can be realized have been suggested~\cite{ring}. 
%Studies of superfluid hydrodynamics in containers with complex geometry
%are obviously relevant for understanding the results of experiments
%on superfluidity in atomic systems confined in such traps. 
Also, there have been many experimental studies 
of the flow properties and NCRI of superfluids confined in porous media
such as vycor glass and containers packed with fine 
powder~\cite{bill,porous2,porous3,porous4}. 
The first experimental observation~\cite{chan3}
of ``supersolid'' behavior was in a torsional oscillator 
experiment on solid $^4{\rm He}$ confined 
in vycor glass. Since the
pores in these systems have complex geometry, it is necessary to work out
the hydrodynamics of superfluids in irregular-shaped channels in order to
understand the results of these experiments in quantitative detail. 
%CDN New lines below %OTVN some editing
%CDNEW deleted the lines below to shorten the discussion of porous media
%Recent experiments~\cite{porous4} have provided strong evidence for the 
%existence of a phase of liquid $^4{\rm He}$ in nanoporous media in which
%islands of superfluid are surrounded by normal liquid, so that there is no
%phase coherence across the sample. Very little information is available at
%present about the geometry of these superfluid islands. Such 
%information can, in principle, be obtained from a combination of accurate
%measurements of the NCRI in this phase and theoretical investigations of
%the hydrodynamics of superfluids in complex confining geometries.  

%OTVNEW  first sentence was repetitive To shed light on the flow properties of 
Thus, we study here the hydrodynamics of a superfluid confined
in a two-dimensional region between two concentric circular arcs, each 
of which subtends an angle $\beta$ at their common center. The annular region
between the two arcs is bounded on two sides by straight walls along
the radial direction. 
Thus, the special case with 
$\beta = 2\pi$ corresponds to a ring that is blocked by a wall placed
perpendicular to its inner and outer peripheries. This two-dimensional geometry
corresponds, neglecting edge effects,
to that used in many experiments on supersolid behavior in 
$^4{\rm He}$ where the helium is
confined in the annular region between two concentric cylinders, under the
assumption that the cylinders are long enough and the confined system is 
homogeneous along the cylinder axis. 
%CDN New line
%CDNEW deleted the line below about cold atomic systems
%Apart from the density variation produced by the presence of a confining
%potential, the cylindrical geometry considered here 
%is also appropriate for cold atomic
%systems confined in cylindrically symmetric magnetic traps such as the
%Ioffe-Pritchard trap~\cite{becv}.
In the limit of vanishing inner radius, %OTVN below removes ambiguous this
the geometry we study corresponds to that of a wedge with opening angle $\beta$. The
limit $\beta = 2\pi$ in this case represents a circular container with a
straight blocking wall extending from the center of the circle to 
its periphery. 

We assume throughout the paper
that the fluid is incompressible, %OTVNEW new sentences below 
which is appropriate for superfluid Helium. %otvrr3 Limitations of this
%assumption for atomic systems are discussed later. 
We first consider the case where  
there are no vortices (so that the superfluid
flow is irrotational), and solve the hydrodynamic equation for the 
velocity field for rotation about an axis perpendicular to the plane of the 
system and passing through the common center of the arcs that form its
boundary. The sample geometry is reflected in the boundary conditions 
for the velocity field.
For incompressible and irrotational flow, the velocity field can be expressed
in terms of either a scalar or a vector potential (stream
function), analogous to those in
electromagnetic theory, both of which satisfy the Laplace equation with 
appropriate boundary conditions. The  scalar potential method is
simpler, and leads to series
that converge rapidly. We have used this method to obtain the 
velocity field for 
$\beta=2\pi$ and $\beta=\pi$. For a general value of $\beta$, however, 
the stream function method, although more difficult in that
%CD replaced vector potential by stream function
it leads to
series that are not convergent, but Borel summable,
is more powerful. We have therefore used
it to obtain the velocity field for arbitrary $\beta$.
We present analytic results for the velocity field and the moment of inertia
for arbitrary values of the inner and outer radii and the opening angle 
$\beta$. We also derive a simple ``parallel axis'' theorem that relates
the moment of inertia for rotation about any axis perpendicular to the plane
of the system to the calculated value for rotation about an axis passing
through the  center of mass.

In the context of experimental observations of NCRI in solid $^4{\rm He}$, 
the most important result of our study is about the NCRI of a blocked ring.
When the ring is blocked, the superfluid can not flow 
through it. However, due to the irrotational nature of
superfluid flow, the moment of inertia is smaller than that for rigid-body
rotation. Therefore, the drop in the moment of inertia 
when the block is removed
(the superfluid does not contribute to the moment of inertia when there is 
no block) is less than the rigid-body value. Our calculations show that the
moment of inertia of a blocked ring whose width is small compared to its
radius is very close to its moment of inertia for rigid rotation, so that 
unblocking the ring (i.e. the opening up of 
a closed path) 
%CD surrounding the rotation axis)
produces a large drop in the moment of inertia (nearly equal to its
rigid-rotation value), which would show up in an experiment as a relatively
large value of the NCRI. 
%CDN New lines below %OTVN edited
Thus, the onset of NCRI in experiments on solid 
$^4{\rm He}$ may correspond to the unblocking of large closed paths in the
network of defects along which the superfluid component is supposed to exist.
Our results for the NCRI of a superfluid 
confined in a blocked ring can be compared directly with those of 
experiments~\cite{chan1,reppyn} in which
the NCRI of solid $^4{\rm He}$ confined in an annular cell is measured 
both in the presence and in the absence of a barrier that blocks flow 
around the annular channel.
We 
show that our results, when combined with accurate measurements of the 
NCRI, can provide valuable information about the
structure of the superfluid network in solid $^4{\rm He}$,
and  discuss the validity of our hydrodynamic description
for superfluid flow in narrow channels such as those along 
crystalline defects in solid $^4{\rm He}$. %OTVNEW small change
%CDNEW Removed reference to He in porous media and added new line about
%applicability of hydrodynamics
% and in liquid
%$^4{\rm He}$ in porous media. 

%OTVN sentence moved.
%CDN Text rewritten below %OTVN some editing
%otvrr3 paragraph below edited and partly moved
An interesting new result of our calculation  
is that the
velocity field  for a wedge with $\beta > \pi$ 
diverges
at the tip of the wedge for {\it any} nonzero value of the angular
velocity $\Omega$. 
This means that the implicit assumption
that the velocity field nowhere exceeds the Landau critical
velocity is in principle
mathematically incorrect for these wedges: % OTV with $\beta > \pi$: 
for any nonzero 
value of $\Omega$, there must be a region near the tip where the liquid 
is in the normal state.  
We show that the size of
the region where this occurs is too small 
%CD changed the wording below . 
to have any measurable consequence in $^4{\rm He}$ experiments  
performed with usual geometries. 
This divergence of the velocity can be removed
by the presence of a single vortex.
We calculate the position of this vortex and
the rotational inertia in its presence.
%CDN2 deleted next sentence because the critical velocity is discussed later
%This result suggests that %otvrr3 vortices may be present in these systems 
%even if the angular speed of
%rotation is arbitrarily small, implying that 
%the critical angular speed for the 
%nucleation of a vortex may be close to zero. 
Our calculations uncover also several %OTVN moved here & edited
interesting mathematical issues and we indicate ways of
addressing them.  Some of these  were also present
in earlier 
studies~\cite{fetter} of superfluid hydrodynamics, while
some are new.  We discuss these questions as they appear throughout the paper.

%CD Changed wording below
Whether vortices appear or not is %OTVN words added & minor edits
in general determined by the 
free-energy cost of creating a vortex.
We will show that for % OTVN parameter values appropriate for 
typical experiments on $^4{\rm He}$,
%the free energy cost of creating a vortex is the relevant one and 
vortices do not occur for sufficiently small angular velocities.
However, as pointed out in Ref.~\cite{fetter}, states with vortices present
will have, at sufficiently larger values of the angular
velocity, a lower free energy than the vortex-free state.
%Their presence at higher speeds is determined from free energy
%considerations. We therefore address here also these considerations, which
We calculate the critical angular velocity for vortex nucleation which
turns out, for typical $^4{\rm He}$ samples, %OTVN
to be in the experimentally important
range of angular velocities. We
show how the rotational inertia is modified by these
vortex excitations. These results are relevant for understanding the 
experimentally observed dependence of the NCRI of ``supersolid'' $^4{\rm He}$
on the frequency of torsional oscillations~\cite{kojima,reppyn,chan4}. 
%otvrr3 rest of paragraph cut In 
%the case of superfluidity in cold atomic systems confined in 
%blocked wedge-shaped traps with $\beta > \pi$, where the  
%velocity in the vortex-free state may exceed the Landau critical velocity in
%a sizeable region near the tip of the wedge, the critical velocity for vortex
%nucleation would be determined from the interplay between the requirements of
%keeping the velocity below the Landau critical value and minimizing the free
%energy. The critical velocity for vortex nucleation may be close %OTVN
%to zero if the free energy cost of creating a normal region dominates over that
%for creating a vortex.  
  
%CD Changed wording below
The rest of this paper is organized as follows. In section \ref{results},
we describe in detail our calculations. We present first
two alternative methods of calculating the
velocity field in the vortex free case,
and discuss the results obtained for this 
field and the moment of inertia. We compare our results for the NCRI with
those of experiments on solid $^4{\rm He}$ in blocked annular geometry and
point out other implications of our results for experimental studies of
superfluidity. %otvrr3 in this and other systems. 
We then explain how to
include vortices in our description and calculate
the critical angular velocity for vortex nucleation. A summary of our results
is presented in the concluding section~\ref{summary}.
  
\section{Results}
\label{results}

%Results of our calculations are described in detail in this section.

\subsection{Formulation of the problem}
\label{form}

We consider, as explained above, superfluid flow in an ideal
cylinder, long enough in the $z$ direction so that edge effects
are negligible and the problem 
quasi two-dimensional. The cross sections of the
cylinders that we will consider will be bounded 
by two concentric circular arcs 
of radii $a$ and $b$ (with $a>b$) and encompassing an angle that we
will call $\beta$. In the limit $b=0$ the shape of this cross section
is that of a circular wedge.
We will consider all values of
$\beta$, $0<\beta\le 2\pi$. It must be emphasized that the case
$\beta=2 \pi$ is not the same as that of a ring, since a boundary along
a radius still exists.

In the absence of vortices (the generalization to the case when vortices are present
will be discussed below) the superfluid velocity field ${\bf v}({\bf r})$ for
an incompressible fluid satisfies:
\begin{subequations}
\label{basic}
\begin{eqnarray}
{\bf \nabla} \cdot {\bf v}({\bf r})=0
\\
%\label{nodiv}
%\end{equation}
%\begin{equation}
{\bf \nabla} \times {\bf v}({\bf r})=0.
%\label{nocurl}
\end{eqnarray}
\end{subequations}
The boundary condition corresponding to superfluid rotation around %otvrr3
some center $O$ %otvrr3 editors reply
%CDN2 changes in wording and the equation
with uniform angular velocity ${\bf \Omega}$ is that~\cite{fetter} the 
normal component of the fluid's velocity at the boundary  must equal 
the normal component of the rigid-body velocity ${\bf \Omega} \times {\bf r}$
at that point. That is,  the component of ${\bf v}({\bf r})$ along 
the outward normal $\hat{\bf n}$ 
to the boundary  must equal, at any point on the boundary, the component 
of ${\bf \Omega} \times {\bf r}$ along $\hat{\bf n}$ at that point:
\begin{equation}
{\bf v}({\bf r})\cdot \hat{\bf n} =({\bf \Omega} \times {\bf r})\cdot \hat{\bf n}
\label{bc}
\end{equation}
were ${\bf r}$ is a vector from $O$ to a point on the boundary. 
%and the index $\perp$ denotes the component perpendicular to the boundary.
The point $O$ is not necessarily the center of mass of the system:
in general we will take it to be, for reasons of
obvious computational convenience, the center of the arc or arcs that are
part of the boundaries of our system.

There are two obvious ways  to solve Eqs.~(\ref{basic}). The first is to introduce
a scalar potential $V({\bf r})$ such that ${\bf v}({\bf r})={\bf \nabla} V({\bf r})$.
In that case  $V({\bf r})$ satisfies the Laplace equation, %otvrr3 inlined
%\begin{equation}
$\nabla^2 V({\bf r})=0$, 
%\label{laplace}
%\end{equation}
and  Eq.~(\ref{bc}) is a Neumann boundary condition on $V$. Alternatively,
one can introduce a stream function  $\Psi({\bf r})$ such that:
\begin{subequations}
\label{vp}
\begin{eqnarray}
v_x=-\partial \Psi/\partial y
\\
v_y=\partial \Psi/\partial x,
\end{eqnarray}
\end{subequations}
where one can think of $\Psi$ as the $z$ component of a vector potential 
$[{\bf v}({\bf r})= -{\bf \nabla} \times (\hat{z} \Psi({\bf r}))]$. It is
obvious that $\Psi({\bf r})$ also satisfies the Laplace equation, %otvrr3
%\begin{equation}
$\nabla^2 \Psi({\bf r})=0$.
%\label{laplace1}
%\end{equation} 
Now, however, the boundary conditions are of the Dirichlet form~\cite{fetter}: at any
point in the boundary,
\begin{equation}
\Psi({\bf r})=\frac{1}{2}\Omega r^2.
\label{bc1}
\end{equation}

It turns out, as we will see, that for certain special values of $\beta$ such as $\pi$ and $2\pi$, the
scalar potential method is much simpler to use and leads
to expressions for ${\bf v}({\bf r})$ in the form of rapidly
convergent series which are very convenient.
However, for other
values of $\beta$, this method becomes rather awkward. The stream function method
on the other hand can be used for any value of $\beta$, but the resulting
expressions  involve asymptotic series. These are, however, Borel summable
and agree with the results obtained from $V({\bf r})$ in the cases
where the scalar potential  method works well. For this reason, we will first present
here results obtained from  $V({\bf r})$ for $\beta=2\pi$ and $\beta=\pi$
and then consider the general case using the stream function.  

Once the velocity field is obtained, the angular momentum (and hence the moment of inertia) can be
calculated by straightforward integration of the velocity field. In this
way, the depletion of the moment of inertia from its rigid body value
is obtained. In general our origin $O$ is not the center of mass (COM)
of the system: therefore it is important to discuss an interesting
property of the nature of
the parallel axis theorem shift in the superfluid case.
If one considers the moment
of inertia of the superfluid with respect to the COM, $I_{SF}^{COM}$ one
finds, of course, that it is always smaller than 
that of the corresponding rigid
object (RO) of the same shape and
density, $I_{RO}^{COM}$. Indeed, for the case of a circle
$I_{SF}^{COM}$ vanishes. With respect to an arbitrary origin $O$ one
has for the superfluid a total moment of inertia 
$I_{SF}^T= I_{SF}^{COM} + I_{SF}^{PA}$ where the last term is the 
parallel axis shift. The key point here is that this shift is the
same as that for the rigid object. One has:
\begin{equation}
I_{SF}^{PA}=I_{RO}^{PA}.
\label{pa}
\end{equation}
The proof of this theorem is very simple: the problem, as defined by the
above equations and boundary conditions, is linear. If one shifts the
origin from the COM to a point a distance ${\bf R}$ away from it,
the velocity field of the boundaries shifts to 
${\bf v}=({\bf r}+{\bf R})\times{\bf \Omega}$. In view of this, the
linearity of the problem,
and the boundary condition Eq.~(\ref{bc}), the solution of the shifted
problem is the velocity field computed with respect to rotations
around the COM, plus a uniform velocity field ${\bf R}\times{\bf \Omega}$.
This second field trivially satisfies the equations and  takes care
of the additional 
term in the boundary condition. But it is trivial to verify that such a 
constant field leads simply to a parallel axis theorem shift in 
the moment of inertia equal to that
for the corresponding rigid object. 
This applies irrespective of the shape of the object: it is not
limited to the wedge shapes considered here. It is straightforward to check
by direct calculation that it applies, for example, to the
ellipsoidal shapes of 
Ref.~\onlinecite{fetter}. This theorem has 
physical
consequences: since the parallel axis shift cannot be ``depleted'' from its
RO value by the superfluid flow, in general the fractional depletion of $I_{SF}$
will always be largest when the rotation is around the COM.  %and conversely,
%rotations around points well away from the COM will be ineffective in
%reducing $I_{SF}$.

\subsection {Scalar potential method for $\beta=2\pi$ and $\beta = \pi$}
\label{scalar}

To illustrate the results, let us first turn to the simplest case
where $\beta=2\pi$, $b=0$ (a circle with a wall along its radius). For
this case, one can very simply use the scalar potential method. We write,
in polar coordinates:
\begin{equation}
V(r,\phi)=\sum_{m \ge 1}a_mr^{m/2}\sin(m\phi/2)+ 
\sum_{m \ge 1}b_mr^{m/2}\cos(m\phi/2).
\label{Vcircle}
\end{equation}
With the radial wall set along the $\phi=0$ direction, the azimuthal
component of the velocity,
\begin{equation}
\begin{split}
v_\phi(r,\phi)&=\sum_{m \ge 1}\frac{m}{2} a_mr^{m/2-1}\cos(m\phi/2) \\ &- 
\sum_{m \ge 1}\frac{m}{2}b_mr^{m/2-1}\sin(m\phi/2)
\end{split}
\label{vtcircle}
\end{equation}
must equal $\Omega r$ at $\phi=0$. This immediately tells us that all
the $a_m$ vanish  except $a_4$, which equals $\Omega/2$. The radial component
is then:
\begin{equation}
v_r(r,\phi)=\Omega r \sin (2\phi)+
\sum_{m \ge 1}\frac{m}{2}b_mr^{m/2-1}\cos(m\phi/2) 
\label{vrcircle}
\end{equation}
At $r=a$ we have $v_r=0$ and %otvrr3 hence:
%\begin{equation}
%-\Omega a \sin(2\phi)= \sum_{m \ge 1}\frac{m}{2}b_ma^{m/2-1}\cos(m\phi/2)
%\end{equation}
from this one obtains that all the $b_n$ with even $n$ are zero
while for odd $n$:
\begin{equation}
b_n=\frac{32 \Omega a}{\pi n (n^2-16) a^{n/2-1}}.
\end{equation}
From these and Eqs.~(\ref{vtcircle}) and (\ref{vrcircle}) we have the final
result for the velocity field:
\begin{widetext}
\begin{subequations}
\label{vcircle}
\begin{eqnarray}
v_r(r,\phi)=\Omega r \sin (2\phi) +
\frac{16 \Omega a}{\pi}
\sum_{n>0,\; \text{ $n$ odd}}\rho^{n/2-1}\frac{1}{n^2-16}\cos(n\phi/2)
\\
v_\phi(r,\phi)=\Omega r \cos (2\phi) -
\frac{16\Omega a}{\pi}
\sum_{n>0,\; \text{ $n$ odd}}\rho^{n/2-1}\frac{1}{n^2-16}\sin (n\phi/2)
\end{eqnarray}
\end{subequations}
\end{widetext}
where $\rho \equiv r/a$.

Two remarks are needed about these simple results: first, %OTVN colon
the series involved are very rapidly convergent.
Second, the velocity components have a square root singularity at the origin.
Mathematically, the %OTVN sentence moved
singularity is integrable, and allows for the formal calculation of the moment
of inertia. 
Physically, %OTVN this is not important: 
the relevant number is the value of $r$ at which
the velocity would exceed  the Landau critical 
velocity $v_c$.
%CDN next paragraph rewritten 
For liquid $^4\rm{He}$, $v_c \approx 2.5 \times 10^4 $cm/s~\cite{huang}, and
%OTV
in typical experiments on supersolid behavior,
the maximum value  
of $\Omega$ 
is less than $0.1 s^{-1}$ (see for example, \cite{kojima,chan1}).    
%or potentially
%up to\cite{chan1} $1000 s^{-1}$. 
This would mean than only at values of $r/a$ 
around $10^{-11}$ would $v_c$ be exceeded. %OTV 
%CD changed wording below
Such small values of $r$ would not have any experimentally
measurable consequence (the hydrodynamic description we use would not even apply
to such length scales). %OTV 
Also this divergence is not present 
for nonzero values of the inner radius $b$, and 
the inner radius is  finite
(of order $10^{-1}$ cm) %OTV
in torsion and
rotation experiments. %OTVN changes below
Thus, this divergence is not important for $^4\rm{He}$.
This divergence may have observable consequences in 
Bose-Einstein condensates (BEC) in cold atomic systems~\cite{bec1,bec2},
%, one would find that the
%region where the velocity exceeds $v_c$ might be a
%sizable portion of the whole system. However %OTVNEW edit  
although our  
incompressibility and uniform density assumptions are not
%conclusions may not be 
%quantitatively %OTVNEW directly
applicable to BEC in cold atomic systems, where the 
high compressibility  
and the confining potential introduces substantial variations in
the density. 
We show later that the divergence discussed above %OTVNEW edit
is present in blocked wedges for all values of $\beta$ greater than $\pi$. 
The effects of this divergence are  
discussed %otvrr3 in detail 
in sections~\ref{vortex} and \ref{freeen}. 

The angular momentum is obtained by integration of $r v_\phi$ over 
the sample and the moment of inertia is just the ratio of the angular 
momentum and the angular velocity $\Omega$. We will use units in
which the areal mass density is unity. We obtain the result:
%CD Here below, I have removed \Omega from the expressions for the 
%moment of inertia.
\begin{equation}
I_{SF}=-\frac{128 a^4}{\pi}\sum_{n>0,\; \text{ $n$ odd}}
\frac{1}{n (n^2-16)(n+4)},
\label{moiold}
\end{equation}
which, after numerically evaluating the rapidly convergent
series, gives $I_{SF}=0.693 a^4$. Thus we have for this obstructed circle: 
\begin{equation}
\frac{I_{SF}}{I_{RO}}\approx 0.441.
\end{equation}

%This result would have to be corrected 
%to include a contribution of a vortex 
%if, as explained above, the experimental situation where such that the formal
%divergence in the velocity field was relevant.

The same method can be used at $\beta=\pi$. In that case the only
significant difference is that in the expression for $V({\bf r})$ one
must write:
\begin{equation}
V(r,\phi)=\sum_{m \ge 1}a_mr^{m}\sin (m\phi)+ 
\sum_{m \ge 1}b_mr^{m}\cos(m\phi).
\label{Vhcircle}
\end{equation}
As before, all the coefficients $a_n$ are determined from the
boundary conditions on $v_\phi$
at $\phi=0$ and $\phi=\pi$. Both are satisfied if 
all $a_n$ vanish except $a_1=\Omega/2$. The $b_n$ are determined then
from the boundary condition on $v_r$. 
The result for the velocity field is:
\begin{widetext}
\begin{subequations}
\label{vhcircle}
\begin{eqnarray}
v_r(r,\phi)=\Omega r \sin (2\phi) +
\frac{8 \Omega a}{\pi}
\sum_{n>0,\; \text{ $n$ odd}}\rho^{n-1}\frac{1}{n^2-4}\cos (n\phi)
\\
v_\phi(r,\phi)=\Omega r \cos (2\phi)-
\frac{8\Omega a}{\pi}
\sum_{n>0,\; \text{ $n$ odd}}\rho^{n-1}\frac{1}{n^2-4}\sin (n\phi).
\end{eqnarray}
\end{subequations}
\end{widetext}
The series are again convergent, and now
the previously found integrable singularity at the origin is absent.
The moment of inertia with respect to the origin is:
\begin{equation}
I_{SF}=-\frac{16 a^4}{\pi}\sum_{n>0,\; \text{ $n$ odd}}\frac{1}{n(n^2-4)(n+2)}.
\end{equation}
Numerically, we have $I_{SF}= 0.488 a^4$ which gives a ratio
$I_{SF}/I_{RO}=0.621$, a value higher than that for the circle. However, we
must recall that in this case $O$ is not the COM and that (as shown above)
there is no reduction in the parallel axis term so that from
the point of view of the COM the reduction must be larger. Indeed one finds that:
\begin{equation}
\frac{I_{SF}^{COM}}{I_{RO}^{COM}}=0.41,
\end{equation}
which is actually a little less than that for the circle.

One can see that it is awkward to extend this simple procedure  to other values of $\beta$. 
If one sets for example $\beta=\pi/2$ and doubles again the angles
and powers in the expression for $V({\bf r})$ one finds that it is not possible to
satisfy the boundary condition for $v_\phi$ at $\phi=0$ and $\phi=\pi/2$
from a single term in the first sum (the $a_n$ coefficients) in the potential.
Similar difficulties are found at e.g. $\beta=3 \pi/2$. Although
these difficulties should not be unsurmountable, we will instead use the
stream function method in the general case and deal appropriately there 
with the mathematical difficulties associated with the asymptotic series that 
then result. 

\begin{figure}
\includegraphics[width=3.2in]{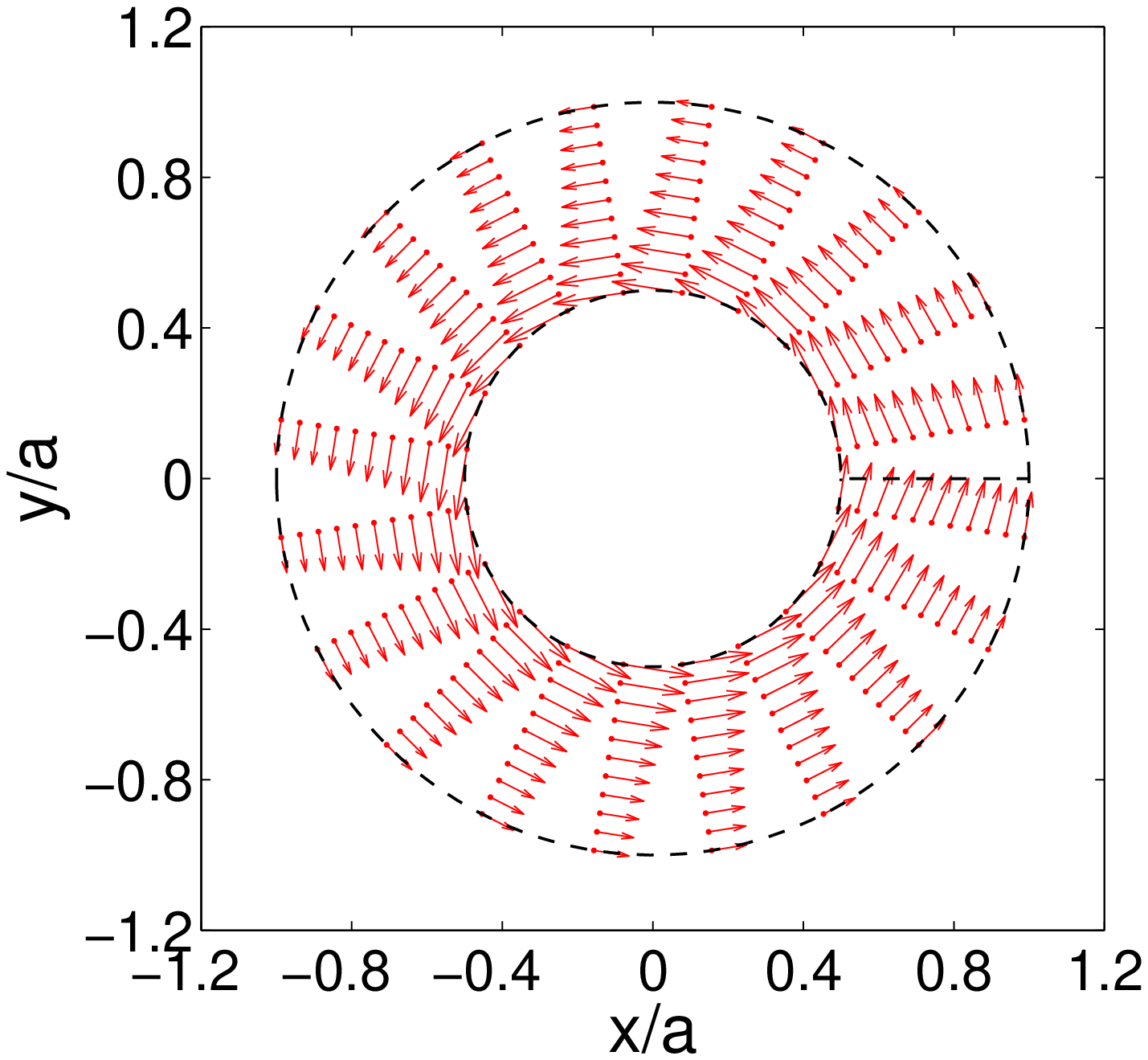}
\includegraphics[width=3in]{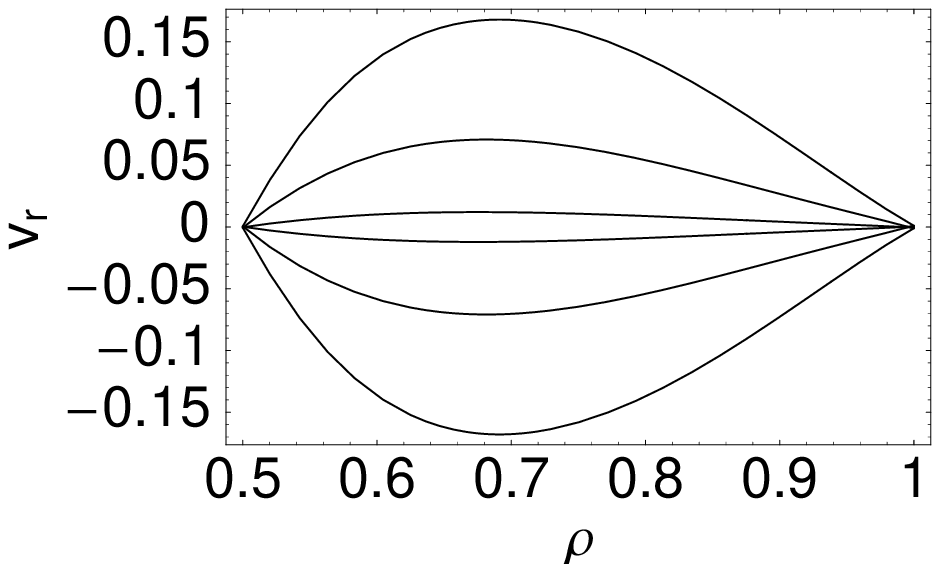}
%\vspace{-2in}
\includegraphics[width=3in]{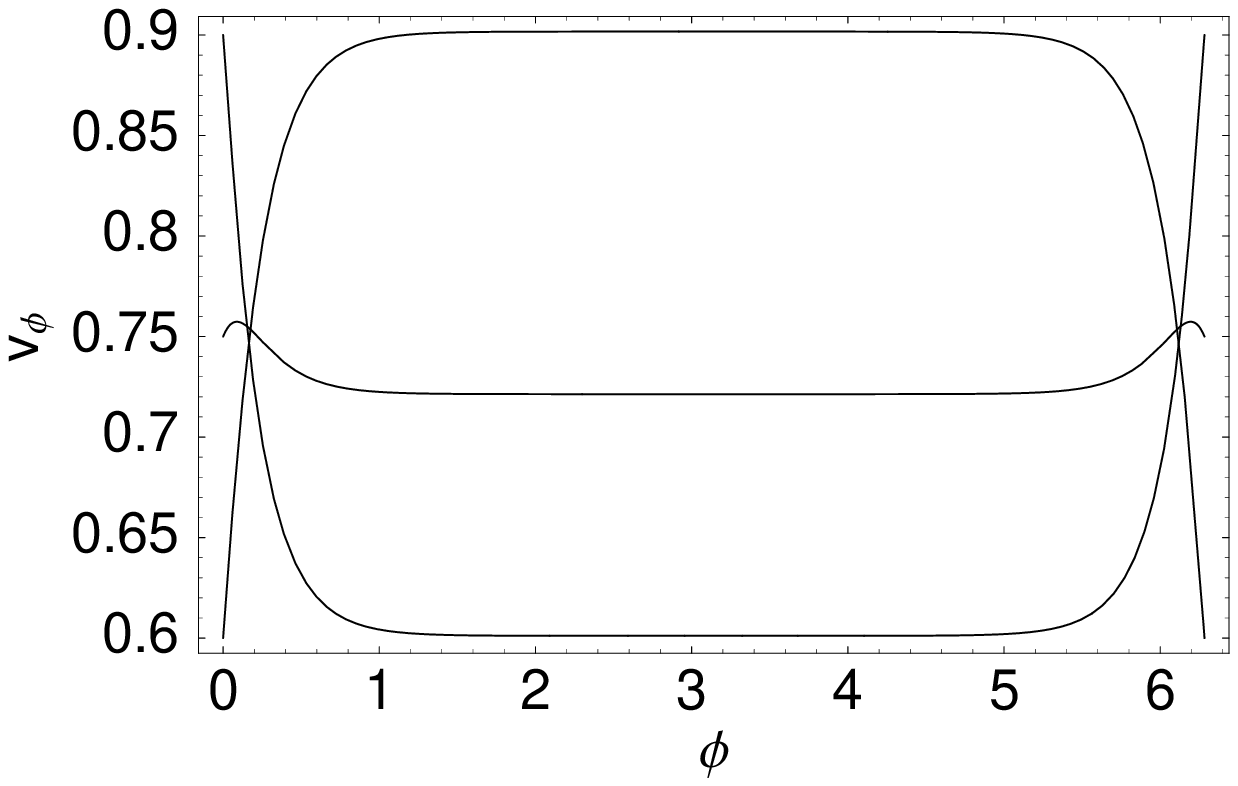}
%\vspace{-0.8in}
\caption{
The velocity field for a blocked ring with $c=0.5$. 
The first panel shows the relative strengths
of the velocity field as a function of position.
The second panel
is the radial component (in units of $\Omega a$) plotted vs $\rho\equiv r/a$
at azimuthal angles $\phi$ (from bottom to top) $\pi/16,
\pi/8, \pi/4, 7\pi/4, 15\pi/8, 31\pi/16$. The third panel, in the same
units, shows
the azimuthal component of the velocity vs. $\phi$ at
$\rho=0.6, 0.75,  0.9$.  }
\label{fig1}
\end{figure}
  
However, one can easily generalize this simple procedure, for the above
values of $\beta$, to the physically
more relevant case where $b>0$.  We will consider here the important
case of an obstructed ring, $\beta= 2\pi$. In that case one simply
has to add to the potential in Eq.~(\ref{Vcircle}) the appropriate negative
powers of $r$. %otvrr3 equation suppressed
%Taking into account directly the boundary condition on
%$v_\phi$ we write:
%\begin{equation}
%V(r,\phi)=\Omega r \sin(2 \phi)+
%\sum_{m \ge 1} \left(b_m r^{m/2}+ \frac{c_m} {r^{m/2}}\right) \cos (m\phi/2).
%\label{Vring}
%\end{equation}  
The coefficients %$b_m$ and $c_m$
are then found from the boundary conditions
on $v_r$ at $r=a$ and $r=b$.  One then obtains the velocity fields:
\begin{widetext}
\begin{subequations}
\label{vring}
\begin{eqnarray}
v_r(r,\phi)=\Omega a \rho \sin (2\phi) %\\ \nonumber &+&
\frac{16 \Omega a}{\pi}
\sum_{n>0,\; \text{ $n$ odd}}\cos(n\phi/2)\frac{1}{(1-c^n)(n^2-16)}
\left[\rho^{n/2-1}f_n(c)-\frac{g_n(c)}{\rho^{n/2+1}}\right],
\end{eqnarray}
\begin{eqnarray}
v_\phi(r,\phi)=\Omega a \rho \cos 2\phi %\\ \nonumber &-&
\frac{16\Omega a}{\pi}
\sum_{n>0,\; \text{ $n$ odd}}\sin (n\phi/2)\frac{1}{(1-c^n)(n^2-16)}
\left[\rho^{n/2-1}f_n(c)+\frac{g_n(c)}{\rho^{n/2+1}}\right].
\end{eqnarray}
\end{subequations}
\end{widetext}
where $c\equiv b/a <1$, $f_n(c)=1-c^{n/2+2}$ and $g_n(c)=c^n-c^{n/2+2}$.
Plots of the fields given by Eqs.~(\ref{vring}) are shown in Fig.~\ref{fig1}.
All the plots in the figure are for $c=0.5$, %OTV
a value in the region where, as we shall see below,
NCRI effects are found to be largest.
In the first panel, the vector
field is displayed in two dimensions over the entire
sample. The units 
of velocity are  arbitrary, but the overall pattern of the field is then
clearly shown.  
In the second and third  panels we 
show a plot of $v_r$ (in units
of $\Omega a$) vs $r$  (in units of $a$) at
several values of the azimuthal angle $\phi$ and a plot, in the same units,
of $v_\phi$ vs $\phi$ at several values of $r$. One can see that the boundary
conditions are satisfied. 

%otvcutpastebegin
\begin{figure}
\includegraphics[width=3.2in]{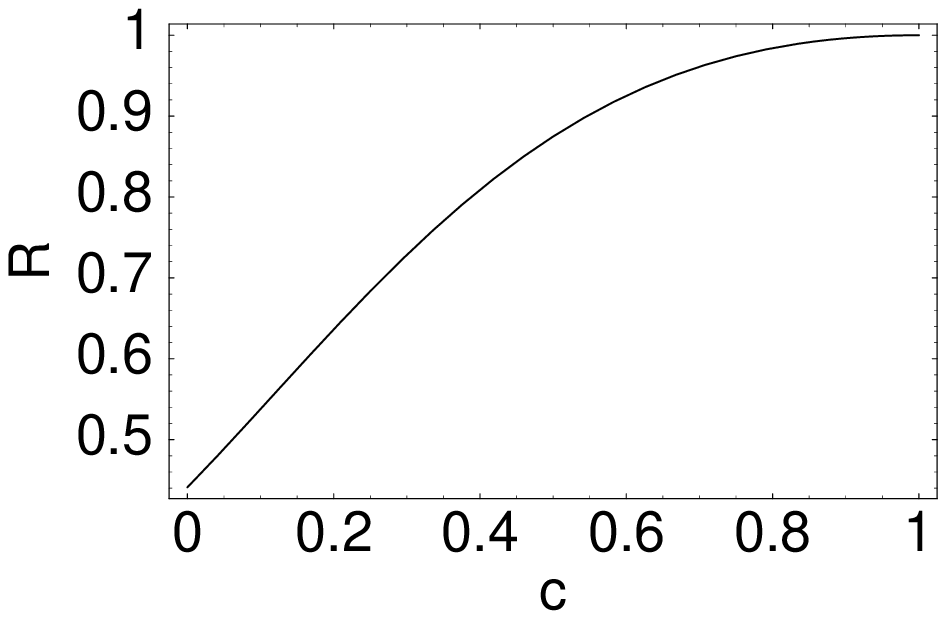}
 %OTV3 filename
\includegraphics[width=3.2in]{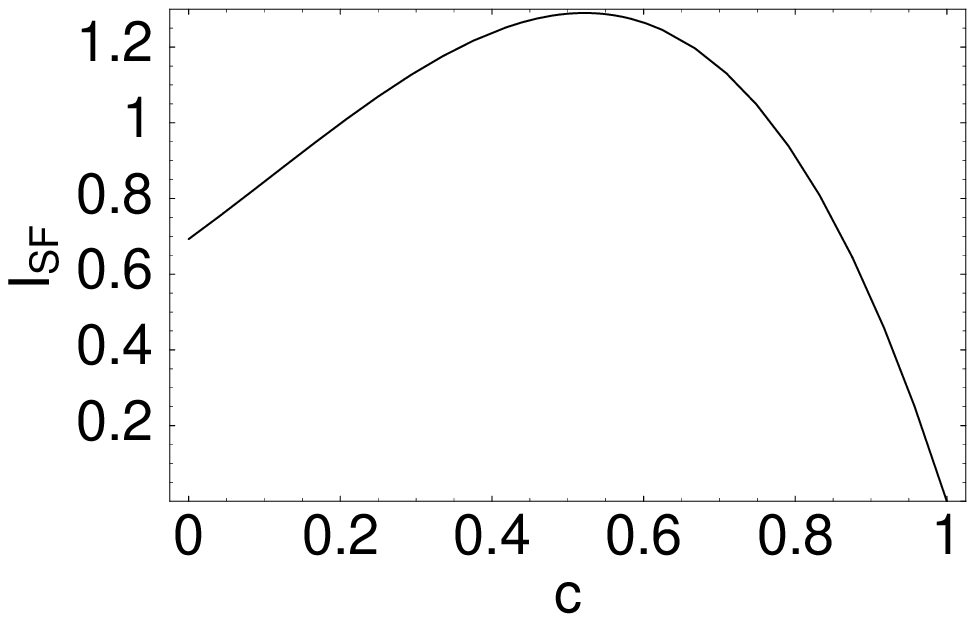}
\caption{Moment of inertia of an obstructed ring in terms
of its aspect ratio $c\equiv b/a$.  In the top
panel the ratio $R$ of $I_{SF}$ (Eq.~(\ref{iring})) 
to the rigid body value is plotted, while in the
bottom panel we plot $I_{SF}$ itself, in units such that $a=1$. The maxima 
in the two plots are at different values of $c$.}
\label{fig2} % caption for the whole figure
\end{figure}
The moment of inertia of the superfluid blocked ring is:
\begin{equation}
\begin{split}
I_{SF}=-\frac{128 a^4}{\pi} &\sum_{n>0,\; \text{ $n$ odd}}
\frac{1}{n (n^2-16)(1-c^n)}\\ & \left[\frac{1}{n+4}f_n^2(c)-\frac{1}{n-4}g_n^2(c)\right],
\end{split}
\label{iring}
\end{equation}
The behavior of this quantity as a function of aspect ratio $c$ is well
worth noting. In the first panel
of Fig.~\ref{fig2} we plot the ratio $R \equiv I_{SF}/I_{RO}$
for a blocked ring of aspect ratio $c$, vs. $c$. As noted above, 
the value for $c=0$ (blocked  circle) would,
strictly speaking, have to be corrected, %to include a contribution 
%from a vortex, 
but the range of $c$ affected by this is
negligible. %The results for all other values of $c$ 
%should be correct if $\Omega$ is sufficiently small. 
The ratio $R$ increases
very quickly with $c$: at $c=1/2$ it already reaches 0.875 while at 
$c=0.75$ it exceeds 97\%. We see, therefore, that  a narrow superfluid
circular channel rotating about its center behaves essentially like a rigid
body when it is blocked. Since, when unblocked, its moment of inertia vanishes,
we see that in such a channel there will be a sharp
discontinuity in $I$ as it is blocked or unblocked. In a sample containing
a number of such channels, discontinuities or glitches in $I$ will occur
as the channels are blocked or unblocked. As $c\rightarrow 1$, $R\rightarrow 1$
and the unblocking would drop R from one to zero, the maximum amount. 
One should recall, however,  that $I$ vanishes at $c=1$ for both the
superfluid and the rigid body. In an experimental situation one would
measure the {\it difference} in $I$ with the channel blocked and unblocked
which  is $I_{SF}$ itself. This quantity has a broad
maximum centered around $c\approx 0.52$ as one can see in the second
panel of Fig.~\ref{fig2}. There we plot $I_{SF}$ itself in units such %OTVN typo
that $a$ is unity. From this plot one can see that the important
experimental contribution would come from a range of rings with 
$c$ values in the region 
0.2 through 0.8.

\subsection{Stream function  method for arbitrary $\beta$}
\label{vector}

As discussed in section~\ref{form}, the velocity field can be written in terms
of a stream function $\Psi({\bf r})$ that satisfies the
Laplace equation with Dirichlet boundary conditions (see Eqs.~(\ref{vp}-
\ref{bc1})). Following Ref.~\cite{fetter}, the general solution for 
$\Psi({\bf r})$ for arbitrary $\beta$ can be written as
\begin{equation}
\Psi({\bf r}) = \frac{1}{2} \Omega \int dl^\prime r^{\prime 2} {\bf{n}}^\prime
\cdot {\bf \nabla}^\prime G({\bf r}^\prime,{\bf r}),
\label{gf1}
\end{equation}
where the line integral $\int dl^\prime$ is over the boundary of the system,
${\bf n}^\prime$ is a unit vector along the outward normal to the boundary,
and $G({\bf r},{\bf r}^\prime)$ is the  Green's function for the 
Laplacian operator, satisfying the equation
\begin{equation}
\nabla^2 G({\bf r},{\bf r}^\prime) = \delta({\bf r}-{\bf r}^\prime),
\label{gf2}
\end{equation}
and the boundary conditions %otvrr3 eqn inlined
%\begin{equation}
$G({\bf r},{\bf r}^\prime) = 0$
%\label{gfbc}
%\end{equation}
for all $\bf r$ on the boundary of the system. Thus, $\Psi({\bf r})$ and
hence, the velocity field, can be obtained from Eq.~(\ref{gf1}) once an 
expression for the Green's function, satisfying Eqs.~(\ref{gf2}) and
its boundary condition %otvrr3 (\ref{gfbc}) 
is obtained.

As in section~\ref{scalar}, we first consider, for simplicity, the case
$b=0$, which corresponds to a wedge of radius $a$ and opening angle $\beta$.
The Green's function in this case is easily obtained~\cite{jackson} to be
\begin{widetext}
\begin{equation}
G(r,\phi;r^\prime,\phi^\prime)= -\frac{1}{\pi} \sum_{n=1}^\infty \frac{1}{n}
r_<^{n\pi/\beta}\left(\frac{1}{r_>^{n\pi/\beta}} - \frac{r_>^{n\pi/\beta}}
{a^{2n\pi/\beta}} \right) \sin(n\pi\phi/\beta) \sin(n\pi\phi^\prime/\beta),
\label{gf3}
\end{equation}
where $r_>$ ($r_<$) is the larger (smaller) one of the two radial coordinates
$r$ and $r^\prime$.
Using this in Eq.~(\ref{gf1}), we obtain the following expression for the 
stream function $\Psi({\bf r})$:
\begin{equation}
\Psi_\Omega(r,\phi) =\frac{2\Omega a^2}{\pi} \sum_{n>0,\; \text{ $n$ odd}}
\sin(n\pi\phi/\beta) \left[\frac{n\pi^2/\beta^2}{n^2 \pi^2/\beta^2 -4}
\left(-\left(\frac{r}{a}\right)^{n\pi/\beta} +\frac{r^2}{a^2} \right)
+\frac{1}{n} \left(\frac{r}{a}\right)^{n\pi/\beta} \right].
\label{psi1}
\end{equation}
The radial and azimuthal components of the velocity field, obtained from
$\Psi_\Omega(r,\phi)$ through Eqs.~\ref{vp} %otvrr3 the relations
%\begin{equation}
%v_r = -\frac{1}{r} \frac{\partial \Psi}{\partial \phi}, \,\,
%v_\phi = \frac{\partial \Psi}{\partial r},
%\label{velocities}
%\end{equation}
are given by
\begin{subequations}
\label{vall1}
\begin{eqnarray}
\label{vr1}
%\begin{equation}
v_r(r,\phi) = \frac{2\Omega a^2}{\pi r} \sum_{n>0,\; \text{ $n$ odd}}
\left( \frac{n\pi}{\beta} \right) \cos(n\pi\phi/\beta)
\times
\left[ \frac{n\pi^2/\beta^2}{n^2 \pi^2/\beta^2 -4} \left(
\left(\frac{r}{a}\right)^{n\pi/\beta} -\frac{r^2}{a^2} \right)
-\frac{1}{n} \left(\frac{r}{a}\right)^{n\pi/\beta} \right],
%\label{vr1}
%\end{equation}
\end{eqnarray}
%and
%\begin{equation}
\begin{eqnarray}
\label{vphi1}
v_\phi(r,\phi) = \frac{2\Omega a^2}{\pi} \sum_{n>0,\; \text{ $n$ odd}}
\sin(n\pi\phi/\beta)
% \\ \nonumber && 
 \times
\left[ \frac{2r}{a^2}\,\, \frac{n\pi^2/\beta^2}{n^2 \pi^2/\beta^2 -4}
-\frac{n\pi}{\beta r} \left(\frac{r}{a}\right)^{n\pi/\beta}
\left( \frac{n\pi^2/\beta^2}{n^2 \pi^2/\beta^2 -4} -\frac{1}{n} 
\right) \right].
%\label{vphi1}
%\end{equation}
\end{eqnarray}
\end{subequations}
\end{widetext}
Calculation of the
velocity field for $\beta = \pi/2$ requires some care because the denominators
of some of the terms in Eqs.~(\ref{vall1}) go to zero for
$\beta = \pi/2$ and $n=1$. The numerators also vanish for these values of 
$\beta$ and $n$, so that finite contributions that vary smoothly with 
$\beta$ across $\pi/2$ are obtained for the velocity components. Similar behavior is found
for $\beta=3\pi/2$ for which the $n=3$ term in the denominators in Eqs.~(\ref{vall1}) vanishes.
These results also exhibit, for $\beta> \pi$, a singularity 
$r^{\pi/\beta-1}$ as $r \to 0$, 
which can be readily  seen from Eqs.~(\ref{vall1}) 
to arise from the $n=1$ term in the sum. 
This is in agreement with what we found
from the scalar potential method. 
%CDN changed text below
As discussed in detail in the
previous subsection, %OTVN typo %this result implies that for larger values of $\beta$,
%the assumption of irrotational
%flow is not valid and a vortex must be present in 
%the system for any nonzero value of $\Omega$. We, therefore, discuss now
%the moment of inertia results only   
%for $\beta \leq \pi$, for which this divergence 
%is not present.
this divergence
is not physically relevant for  $^4\rm{He}$, but may have %OTVN edit
observable consequences in experiments on cold atomic systems.
Its possible physical effects  are discussed in %OTVN
sections~\ref{vortex} and \ref{freeen}.
This singularity is always
integrable: %OTVN 
therefore, 
the angular momentum 
of the superfluid about the origin (tip of the wedge) is easily calculated
for all $\beta$ 
using these expressions for the velocity components.
The result for the moment of inertia about $O$ is
\begin{equation}
I_{SF} = \frac{2a^4}{\pi} \sum_{n>0,\; \text{ $n$ odd}} \frac{1}{n} \left(
\frac{n\pi}{\beta}+4\right)\frac{1}{(n\pi/\beta+2)^2}.
\label{moi1}
\end{equation}
For the case $\beta=2\pi$, the moment of inertia about the origin is given
by the infinite series
\begin{equation}
I_{SF}(\beta=2\pi) = \frac{4a^4}{\pi} \sum_{n>0,\; \text{ $n$ odd}} 
\frac{1}{n} \frac{n+8}
{(n+4)^2}.
\label{moi2}
\end{equation}
This infinite series appears to be different from the one in Eq.~(\ref{moiold})
which was obtained using the scalar potential method. In particular, the 
series in Eq.~(\ref{moi2}) converges more slowly than the one in 
Eq.~(\ref{moiold}). However, %otvrr3 using the identity
%\begin{equation}
%\sum_{n>0,\; \text{ $n$ odd}} \frac{1}{(n-4)(n+4)} = 0,
%\label{ident}
%\end{equation}
it can easily be shown that these two expressions for the moment of inertia are
mathematically identical. We have also checked that a similar
situation applies when the results  for the moment of inertia
obtained from Eqs.~(\ref{vall1}) for $\beta=\pi$ are compared
to those obtained in the preceding section using the scalar potential
method.

However, the situation is much more complicated when, instead of comparing
the moments of inertia,  one compares directly the velocity fields obtained
by the two methods. In this case
it is not sufficient to add or subtract
a series that converges to zero. The reason is that while the series
in Eqs.~(\ref{vcircle}) converge for all 
angles $\phi$ and for any $r \neq 0$, those
in Eqs.~(\ref{vall1}) and (\ref{psi1}) do not. %OTV 
%CD deleted a few lines here
%This lack of
%convergence occurs already in the result for $\Psi_\Omega$, Eq.~(\ref{psi1}), and
%is not limited to $\beta=2 \pi$ but it occurs at all
%angles. It is thus unrelated to the small $r$
%singularity and reflects  a general 
%technical problem with the stream function method: Eqs.~(\ref{psi1}) and
%(\ref{vall1}) involve nonconvergent 
%asymptotic series.  
This question is related to
other technical difficulties with the result (\ref{psi1}), and in
%OTV
general with the stream function method, which we will further address below.

\begin{figure}
\includegraphics[width=3.2in]{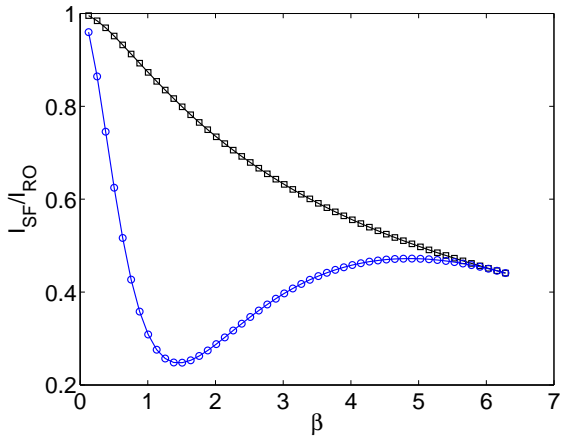}
\caption{The ratios $I_{SF}/I_{RO}$ (upper curve),
and $I_{SF}^{COM}/I_{RO}^{COM}$ (lower curve)
for a superfluid wedge
as a function of the opening angle $\beta$, $0 < \beta \leq 2 \pi$.
$I_{SF}$ is calculated from Eq.~(\ref{moi1}).}
\label{fig3} % caption for the whole figure
\end{figure}

The moment of inertia of the wedge for rigid-body rotation about $O$ is
$I_{RO} = \beta a^4/4$, and its moment of inertia for rigid-body rotation
about its COM
is given by $I_{RO}^{COM} = I_{RO} - I_{RO}^{PA}$ with $I_{RO}^{PA} =
8a^4 \sin^2(\beta/2)/(9\beta)$. Using these results and Eq.~(\ref{moi1}),
we have calculated the ratios $I_{SF}/I_{RO}$ and $I_{SF}^{COM}/I_{RO}^{COM}$
as functions of the angle $\beta$. The results are shown in Fig.~\ref{fig3}.
These ratios are of course less than unity, the level of suppression being
given by the NCRI effect. 
In the figure we see that this fractional suppression is always larger in the $COM$
frame, that is, 
$I_{SF}/I_{RO}$ is always
higher than $I_{SF}^{COM}/I_{RO}^{COM}$, 
except of course at
$\beta=2\pi$ where the two are the same. This is in
agreement with the theorem proved at the end of Sec.~\ref{form}.  
It is interesting that the ratio
$I_{SF}^{COM}/I_{RO}^{COM}$ is not a monotonic function of $\beta$ -- it 
exhibits a minimum at $\beta=\pi/2$.

\begin{figure}
\includegraphics[width=3.2in]{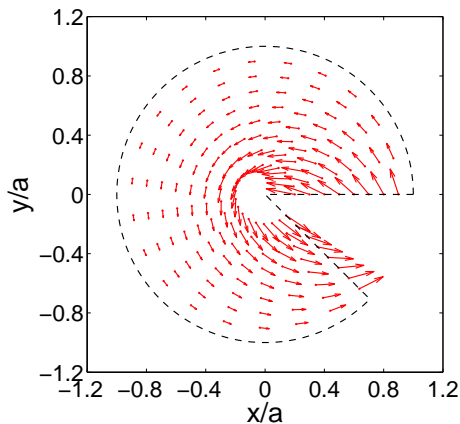}
\caption{
Plots of the velocity field inside the wedge for  
$\beta=(7/8)2\pi$. This should be compared with the first panel  %OTV
of Fig.~\ref{fig1}.
}
\label{fig4}
\end{figure}

A representative plot of the velocity field for  a wedge
with $\beta=(7/8) 2\pi$ 
is shown in Fig.~\ref{fig4}. The velocity vector field is plotted in
arbitrary relative units, as in the first panel of Fig.~\ref{fig1}. It
is instructive to compare that panel with Fig.~\ref{fig4}. In the earlier
case we have $c=0.5$ whereas in Fig.~\ref{fig4} we have a wedge, $c=0$.
The rise in the absolute value of the velocity as $r\rightarrow 0$ can now
be seen. On the other hand, the behavior of  of $v_r$ as a function of 
$\phi$ is clearly very similar: it follows from the second 
panel of Fig.~\ref{fig1}
that $v_r$  is very small except for angles near the radial boundaries, and this 
is clearly the case also for this $c=0$ wedge. The behavior of $v_\phi$
with $\phi$ is also quite similar.   

We now return to the technical difficulties with the 
general solution for the velocity field obtained above
via the stream function. As noted in 
section~\ref{form}, the quantity $\Psi(r,\phi)$ should be equal to 
$\Omega r^2/2$ at all points on the boundary, and the physical velocity
field should satisfy the boundary conditions $v_\phi (r,\phi) = r \Omega$ 
for $\phi=0,\,\beta$ and $v_r (r,\phi)= 0$ for $r=a$. It is easily seen
from Eqs.~(\ref{psi1}) and (\ref{vphi1}) that both $\Psi(r,\phi)$
and $v_\phi (r,\phi)$ vanish for $\phi=0$ and $\phi=\beta$ (since
$\sin(n\pi \phi/\beta) = 0$ for these values of $\phi$). Thus the
boundary condition on the radii appears to be violated even though the 
construction of the vector potential via the Green's function would
seem to ensure that it will not be. As to
Eq.~(\ref{vr1})
for the radial component of the velocity, it can be written as
\begin{widetext}
\begin{eqnarray}
v_r(r,\phi) = \frac{8\Omega a^2}{\beta r} \sum_{n>0,\; \text{ $n$ odd}}
\cos(n\pi\phi/\beta) \frac{1}{n^2 \pi^2/\beta^2 -4} \left[
\left(\frac{r}{a}\right)^{n\pi/\beta} -\frac{r^2}{a^2} \right] %\nonumber \\
- \frac{2\Omega r}{\beta}
\sum_{n>0,\; \text{ $n$ odd}} \cos(n\pi\phi/\beta).
\label{vr2}
\end{eqnarray}
\end{widetext}
While the first term on the right-hand side of this equation vanishes for
$r=a$, the second term does not. Thus, this component
also  appears not to satisfy the required
boundary conditions. 
Numerically, however, we have found that these quantities do
approach  values consistent with the required boundary conditions as the
boundaries are approached from inside, but there is a 
discontinuity as the boundary is approached and
the values exactly at the 
boundaries do not satisfy the boundary conditions. This does not affect the
calculated values of the angular momentum and the moment of inertia because
these quantities are not sensitive to the values of the velocity components
exactly at the boundary.  

However, this numerical argument is not fully satisfactory.
Fortunately
there are better ones. First, one can see that this behavior is
associated with the nonconvergence of the series. The last term
in Eq.~(\ref{vr2}), for example,  is not merely nonzero: the series
that it contains is not convergent while that in the first term is. Indeed
the rearrangement of terms leading from Eq.~(\ref{vr1}) to Eq.~(\ref{vr2})
isolates just this nonconvergent part. However, by rewriting the cosines
in terms of exponentials one can verify that the series in the
last term of Eq.~(\ref{vr2}) is Borel 
summable\cite{BO} (and also Euler summable) with the result being zero. 
With this proviso, Eq.~(\ref{vr2}) satisfies the boundary
condition analytically. Similar arguments can be made for $\Psi_\Omega$
and for the azimuthal component of the velocity.
 
This mathematical problem can 
also be solved by re-defining
the stream function as
\begin{equation}
\Psi(r,\phi) \rightarrow  \Psi(r,\phi)-\frac{2 \Omega r^2}
{\pi} \left[ \sum_{n>0,\; \text{ $n$ odd}} \frac{1}{n} \sin(n\pi\phi/\beta)
-\frac{\pi}{4} \right],
\end{equation}
where the first term in the right side
is that given by Eq.~(\ref{psi1}).
The second term in the right
side, which is subtracted from the old expression, is zero for all points
inside the wedge~\cite{gr}, and is equal 
to $-\Omega r^2/2$ for $\phi = 0,\,\beta$.
Therefore, the subtraction of this quantity does not affect the behavior of
$\Psi(r,\phi)$ inside the wedge (where it still satisfies the Laplace 
equation). At the same time, the redefined $\Psi(r,\phi)$ satisfies the
required boundary condition for $\phi = 0,\,\beta$.
The new term leads to the following additional terms in $v_\phi$ and $v_r$:
\begin{equation}
v_\phi(r,\phi)\rightarrow v_\phi(r,\phi) - \frac{4 \Omega r}
{\pi} \left[ \sum_{n>0,\; \text{ $n$ odd}} \frac{1}{n} \sin(n\pi\phi/\beta)
-\frac{\pi}{4} \right],
\end{equation}
where again the first term in the right side
is the previous result, in this case Eq~(\ref{vphi1}).
The added quantity is zero at all points inside the wedge, and is equal to
$\Omega r$ for $\phi = 0,\,\beta$, so that the required boundary conditions
for these values of $\phi$ are now satisfied. The equation for $v_r$ becomes
\begin{equation}
v_r(r,\phi) \rightarrow v_r(r,\phi) + \frac{2\Omega r}{\beta}
\sum_{n>0,\; \text{ $n$ odd}} \cos(n\pi\phi/\beta).
\label{vr3}
\end{equation}
The new term, added to Eq.~(\ref{vr1}),
cancels the ``offending'' second term in Eq.~(\ref{vr2}), so that
the re-defined $v_r$ satisfies the required boundary condition at $r=a$.

A similar problem with boundary conditions is also present 
in the solution given
in Ref.~\cite{fetter} for the velocity field inside a cylinder with a 
rectangular cross section. The expression for the stream function given in
Eq.~(62) of Ref.~\cite{fetter} does not 
in fact satisfy the required boundary conditions
 posed there
at all points on the boundary. As in the case
considered here, this does not affect the 
results for the calculated physical quantities
in Ref.~\cite{fetter}, and this  mathematical
problem can be cured by the addition of a term similar to the one 
considered above.

\begin{figure}
\includegraphics[width=3.2in]{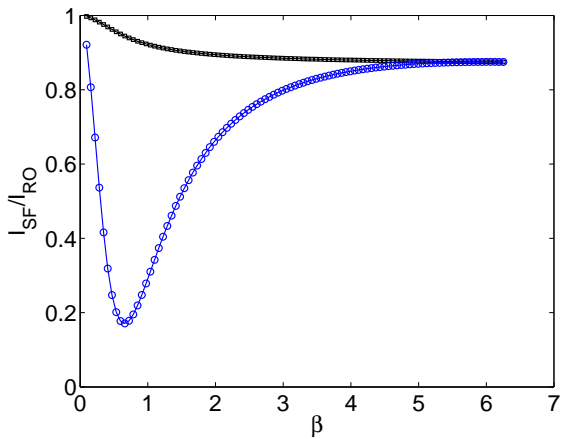}
\caption{The ratios $I_{SF}/I_{RO}$ (upper curve),
and $I_{SF}^{COM}/I_{RO}^{COM}$ (lower curve)
for an annular wedge (Eq.~(\ref{moi3})) plotted 
as a function of the opening angle $\beta$, $0 < \beta \leq 2 \pi$,
at a fixed value of $c=0.5$}
\label{fig5}
\end{figure}

The above calculations can be modified readily to treat a superfluid confined
in the annular region between two concentric arcs with radii $a$ and $b$
($a > b$). The Green's function in this case has the form
\begin{widetext}
\begin{eqnarray}
\label{grfn}
G(r,\phi;r^\prime,\phi^\prime)= -\frac{1}{\pi} \sum_{n=1}^\infty 
\frac{1}{n} \frac{1}{1-(b/a)^{2n\pi/\beta}}
\left( r_<^{n\pi/\beta} - \frac{b^{2n\pi/\beta}}{r_<^{n\pi/\beta}} \right) %\\
%\nonumber
\times \left(\frac{1}{r_>^{n\pi/\beta}} - \frac{r_>^{n\pi/\beta}}
{a^{2n\pi/\beta}} \right) \sin(n\pi\phi/\beta) \sin(n\pi\phi^\prime/\beta).
\end{eqnarray} 
In this case one does not have to worry about the behavior as $r \to 0$.
Asymptotic series in the summations over $n$ are again
encountered and handled as in the
preceding case.
Using this in Eq.~(\ref{gf1}), the stream function $\Psi(r,\phi)$, and from 
it, the radial and tangential components of the velocity are obtained. We 
skip the long expressions for these quantities and quote the final result
for the moment of inertia about the origin:
\begin{eqnarray}
I_{SF} = I_{RO} -\frac{16 a^4}{\beta} \sum_{n>0,\; \text{ $n$ odd}}
\frac{1}{x_n^2(x_n^2-4)} \left[
\frac{x_n^2+4}{2(x_n^2-4)}(1-c^4) \right. %\nonumber \\
- \left. \frac{2x_n}{x_n^2-4} \frac{1}{1-c^{2x_n}} \{(1+c^4)(1+c^{2x_n})
-4 c^2 c^{x_n} \}\right ].
\label{moi3}
\end{eqnarray}
\end{widetext}
Here, $x_n=n\pi/\beta$, $c=b/a$, and $I_{RO} = \beta (a^4-b^4)/4$ is the moment
of inertia for rigid-body rotation. We have checked that this expression 
reduces to that in Eq.~(\ref{moi1}) for $b=0$, and to that in Eq.~(\ref{iring})
for $\beta = 2\pi$.
 In Fig.~\ref{fig5}, we show results for the NCRI in
an annular wedge, as obtained from Eq.~(\ref{moi3}). The plots are the
same as in Fig.~\ref{fig3} except that now we have $c=0.5$, in other words, the
fields are as in Fig.~\ref{fig1}. Again, the fractional suppression is larger, as it must
%CDN2 minor change in wording below
be, in the $COM$ and it exhibits a maximum as a function of $\beta$.

%CDN New pargraphs below
%OTVN numerous edits, tried to tighten
%OTVN Query: separate subsection \subsection {Experiments with artificially
%OTVN blocked wedges}
%OTVN \label{wedges}
The results derived above have a direct relevance to torsional
oscillator experiments on 
solid $^4{\rm He}$~\cite{chan1,reppyn} in which the helium is confined in the
annular region between two concentric cylinders and the NCRI is measured both
in the presence and in the absence of a barrier that prevents flow around the
annulus. If the NCRI in the absence of the barrier 
is due to superflow along a closed
channel surrounding the rotation axis (the 
common axis of the inner and outer cylinders), then the measured value of the
NCRI when the barrier is not present should be 
$(\Delta I)_{\rm{open}} = \rho_s I_{RO}$
where $\rho_s$ is the supersolid fraction and $I_{RO}$  the rigid-body moment
of inertia of the channel of flow about the rotation axis. 
The NCRI in the presence of the barrier should be given by 
$(\Delta I)_{\rm{closed}} = \rho_s (I_{RO}-I_{SF})$ 
where, if this channel is approximately
circular,  $I_{SF}$ is the moment of 
inertia of a blocked superfluid ring calculated above. Thus, the ratio
$R^\prime \equiv (\Delta I)_{\rm{closed}}/(\Delta I)_{\rm{open}}$ 
should be equal to $(I_{RO}-I_{SF})/I_{RO} = 1-R$, where $R \equiv I_{SF}/I_{RO}$
depends (see Eq.~(\ref{iring}) and Fig.~\ref{fig2}) 
on the value of $c=b/a$. % where $a$ and $b$
%are the outer and inner radii of the channel, respectively, and $t \equiv a-b$
%is its thickness.
If the superfluid component were distributed 
homogeneously throughout the sample, then $a$ and 
$b$ would be the outer and inner radii of the annular cell. 
Whether this is the case can be determined by
comparing the experimentally 
measured value of $R^\prime$ with $(1-R_0)$, where  
$R_0$ is the value of $R$ obtained
from Eq.~(\ref{iring}) using these values of $a$ and $b$. 
If the superfluid
is instead confined in a channel (or in several separate 
channels) with width substantially smaller than that
of the annular cell, then $R^\prime$ should be smaller than $(1-R_0)$ because
$R$ increases as the width of the ring is decreased (see 
Fig.~\ref{fig2}).  

In the experiment of Ref.~\cite{chan1}, $a$ = 0.75 cm and $b$ = 0.64 cm, so that
%$c$ = 0.853 for which $R_0$ = 0.99183 and 
$(1-R_0)$ = 0.00817. The experimental
value of $R^\prime$ is 0.015 which is within a factor of two of $(1-R_0)$ but, 
surprisingly, it is higher.
However, the value of $(\Delta I)_{\rm{open}}$ 
appropriate for the
blocked cell was evaluated 
%(from a hydrodynamics calculation) %OTVNEW! 
%CDNN Deleted the text you added because the results being referred to were experimentally
%obtained, not from a calculation
from the  results of a 
different experiment using another 
cell, so that
the quoted value of $R^\prime$ may not be very accurate. % because
% in the calculation of $R^\prime$, . %OTVN moved
Also, a value of $R^\prime$ larger than $(1-R_0)$ may be rationalized by assuming that
the sample contains a large number of narrow superfluid channels, most of which
do not form  closed paths around the annulus (i.e. have $\beta < 2\pi$). These
``naturally blocked'' channels make small contributions to the net sample NCRI.
These contributions are not strongly affected by the imposition of
the external barrier, which can change the value of $\beta$ for
the channels it intersects: our calculation shows that  $R = 
I_{SF}/I_{RO}$ for narrow annular wedges with $\beta < 2\pi$ 
is rather insensitive to
$\beta$. Since these channels contribute almost equally to
$(\Delta I)_{\rm{open}}$ and  $(\Delta I)_{\rm{closed}}$, 
the value of the ratio $R^\prime$ would increase. 

More recently, both  $(\Delta I)_{\rm{open}}$ and  
$(\Delta I)_{\rm{closed}}$ have been measured using the same 
cell~\cite{reppyn}. In this experiment, two cells, both with $a$ = 0.794 cm 
and $b$ = 0.787  and 0.745 cm %$\mu$m and 487 $\mu$m, 
were used. In both cases 
the NCRI in the blocked configuration was found to be smaller than the 
resolution of the experiment.
This is consistent with our calculated
values of $(1-R_0)$  which are 2.9$\times 10^{-5}$ and 
1.3$\times 10^{-3}$, respectively. 
%The upper limit of the value of
%$R^\prime$ was estimated to be 0.047 and 0.08, respectively, for the two cells.
%The small values of $R^\prime$ obtained in these experiments confirm that the
%observed NCRI is due to superflow around the annulus. %OTVNEW edit and merge
%with beginning of next para 
Although the measurements
are not sufficiently accurate to provide more detailed information about the
channels of superflow, %Is is surprising that the measured values of $R^\prime$
%tend to be larger than $(1-R_0)$: as noted above, $R^\prime$ should be smaller
%than $(1-R_0)$ if the superfluid channels are narrower than the annulus.
%CDNEW New lines about the validity of hydrodynamics for narrow channels
it is clear  that more accurate measurements
of $R^\prime$ for samples with different $a$ and $b$, combined with 
the results of our calculations, would be very useful for
elucidating the geometry of superfluid channels in solid $^4{\rm He}$. 

%OTVNEW different paragraph start 
If the superfluid channels are very narrow,  the validity of the 
hydrodynamic description used here (and elsewhere~\cite{chan1}) might %OTVNEW
be questioned. %OTVNEW edits below
However, recent numerical 
studies~\cite{prokofiev1,prokofiev2}  indicate that 
%superfluid channels along crystalline defects have
%lateral dimensions of atomic scale: 
the diameter of the 
superfluid region near the core of a dislocation and
%CDNN: I don't know why you deleted the part about dislocations %OTVNN added again
the width of the 
superfluid layer along a grain boundary are of the order of 
a few nanometers
($\sim 10$ interparticle spacings). %OTVNN %OTVNEW  and many edits and moves below
%CDNN: \gtrsim 10 is probably not true -- the values I saw in the paper are closer
%to 5 interparticle spacings -- that's why I wrote 5-10 %OTVNN changed to 10
These values of the %OTVNN restored your order
superfluid layer width are likely to be lower bounds, %OTVNN restored your order
since superfluid channels of such small lateral dimensions can not explain the
relatively large
superfluid density measured in recent torsional oscillation 
experiments~\cite{reppyn}. 
It has been suggested~\cite{balibarn} that the effective lateral
dimension of the
superfluid region near a crystalline defect may be larger due to a kind of
``proximity effect'', as in superconductors. Also,
studies~\cite{sc} of the
thermodynamics of a system of
interacting vortex lines in type-II superconductors, which can be mapped to the
zero temperature quantum mechanics of a two-dimensional system of 
interacting bosons, show that the width of grain boundaries can exceed 
15-20 interparticle spacings in some cases. 
A hydrodynamic
description should be valid if the width of the typical superfluid regions
is of order $\sim$ 10 interparticle spacings or more: this has %OTVNN
been well-established quantitatively in several numerical 
studies of the flow properties of classical liquids
through narrow channels~\cite{hydro1,hydro2}. % OTVNEW cut repeat have established that a
%hydrodynamic description leads to quantitatively accurate results if the
%channel diameter exceeds about 10 interparticle spacing. 
The same should
to be true for superfluid $^4{\rm He}$ because its coherence length is very
small.

%OTVNEW 
A related effect that needs to be considered if the superfluid channel along
a crystal defect is very narrow is the modulation of the density of the 
superfluid due to the potential arising from the surrounding crystalline
region. We expect our calculations to be valid in the presence of such 
density modulations because the hydrodynamic equation for a rotating 
superfluid 
derived (for low angular speed) 
in a recent study~\cite{josserand} in which superfluidity is assumed
to coexist with a periodic modulation of the density (Eq.~(8) of 
Ref.~\cite{josserand}) is identical to that used in our calculation. %OTVNEW

%Such
%measurements for $^4\rm{He}$ in nanoporous media would also be helpful for
%obtaining detailed information about the superfluid islands that are believed
%to exist~\cite{porous4} in these systems.   

%CD changed heading of subsection
\subsection {Formation of vortices in a wedge with $\beta > \pi$}
\label{vortex}

As noted above, the velocity field obtained from a calculation
in which it is assumed to be irrotational exhibits a 
divergence as $r \to 0$ for a wedge with $\beta > \pi$. 
Thus $v_c$  
must be exceeded near $r=0$, implying that either there is a region
of normal fluid near the tip of the wedge, 
or a vortex is present in the system. 
As we have indicated, %OTV
this issue is unimportant in the torsional oscillation
experiments because the region of normal fluid near the tip would be 
unobservably small for 
experimentally relevant parameter values. 
%CDN Changed wording below
%CDNN: changed therefore to however in the line below
It is, however, 
interesting to inquire 
about the behavior in the general case.
We show here that this divergence in the velocity field is eliminated by the
introduction of a single vortex.  
%CDN2 removed reference to BEC, as demanded by referee 3
%Therefore, this effect may have
%observable consequences in experiments on Bose-Einstein condensates in cold
%atomic systems, in that one would observe the appearance of vortices no
%matter how small the angular speed. 

From symmetry,  the vortex must be located along 
the line $\phi=\beta/2$.
Let the position of the vortex be $(r_v,\beta/2)$. The presence 
of a vortex
of circulation $\kappa (= h/m$, where $h$ is Planck's constant and $m$ is the
mass of a particle of the fluid) at $(r^\prime,\phi^\prime)$ 
leads to an additional term,
$\kappa G(r,\phi;r^\prime,\phi^\prime)$ in the expression for the stream
function $\Psi(r,\phi)$ where $G(r,\phi;r^\prime,
\phi^\prime)$ is the Green's function given in Eq.~(\ref{gf3})
(see Section 3 of Ref.~\cite{fetter} for a derivation
of this result).
This additional term in $\Psi(r,\phi)$ (with $r^\prime
= r_v, \phi^\prime=\beta/2$) leads to the
following additional term in the expression for the radial component of the
velocity near $r=0$:
\begin{widetext}
\begin{equation}
v_r(r,\phi) = v_r^0(r,\phi) + \frac{\kappa}{\beta r}
\sum_{n=1}^\infty r^{n\pi/\beta}
\left(\frac{1}{r_v^{n\pi/\beta}} - \frac{r_v^{n\pi/\beta}}
{a^{2n\pi/\beta}} \right) \cos(n\pi\phi/\beta) \sin(n\pi/2) \equiv v_r^0+v_r^1,
\label{vr4}
\end{equation}
\end{widetext}
where $v_r^0(r,\phi)$ is the curl-free result as given by Eqs.~(\ref{vr3}).
The $n=1$ part of the additional term cancels the divergent $n=1$ contribution
of the previous expression if
\begin{equation}
\kappa \left( \frac{1}{r_v^{\pi/\beta}}- \frac{r_v^{\pi/\beta}}
{a^{2\pi/\beta}} \right) = 8 \Omega a^{2-\pi/\beta} \frac{1}{4-\pi^2/\beta^2}.
\label{vcond}
\end{equation}
It is easy to check that the divergence in the expression for the azimuthal
component of the velocity is also removed if this condition is satisfied.
%Equation (\ref{vcond}) has to be solved to determine the vortex position $r_v$.
%This equation can be written as
%\begin{equation}
%\frac{\kappa}{8 \Omega a^2} (4-\pi^2/\beta^2) \left[ \left(\frac{a}{r_v}
%\right)^{\pi/\beta} - \left(\frac{r_v}{a}\right)^{\pi/\beta} \right] =1.
%\label{vorpos}
%\end{equation}
Defining $(r_v/a)^{\pi/\beta} \equiv \xi$, %is equivalent to
%\begin{equation}
%\frac{1}{x}-x=c, \,\,\hbox{with} \,\, c \equiv \frac{8 \Omega a^2}{\kappa (4-
%\pi^2/\beta^2)}> 0.
%\label{vorpos1}
%\end{equation}
the solution of Eq.~(\ref{vcond}) is $\xi=[\sqrt{4+\eta^2}-\eta]/2$, 
where,
\begin{equation}
\eta \equiv \frac{8 \Omega a^2}{\kappa (4-
\pi^2/\beta^2)}> 0.
\end{equation}
One sees that $\xi$ has the
nice property that $0< \xi < 1$ for any value of $\eta$. The value of $\xi$
changes from 1 to 0 as the dimensionless parameter 
$\gamma \equiv \Omega a^2/\kappa$
increases from zero to a large value, i.e. the vortex moves inward from the rim of
the wedge to its tip as the angular velocity increases.  
%For 
%$\Omega a^2 \approx \kappa$, the vortex would sit near the middle of the wedge.
%This shows that for any
%value of $\Omega$ and the lowest allowed value of $\kappa$ ($=h/m$),
%the divergence in the velocity can be removed by placing a vortex at a point
%$r_v$ that lies between 0 and $a$. For small $\Omega$, $c$ is small and the
%value of $x$ is close to 1, i.e. the vortex is close to the rim, $r=a$. 
%For large
%$\Omega$, $c$ is large and the vortex sits close to $r=0$. Thus, the vortex
%moves inward as $\Omega$ is increased.

Using the expressions for the radial and tangential components of the velocity
in the presence of a vortex, the total angular momentum of the superfluid can
be calculated. The presence of the vortex increases the angular momentum 
about the origin by the amount $L_v$
%\begin{equation}
%L_V = \frac{8\kappa}{\pi} \sum_{n>0,\; \text{ $n$ odd}} (-1)^{\frac{n+1}{2}} \frac{1}
%{n(4-n^2\pi^2 /\beta^2)} \left[\left(\frac{r_v}{a}\right)^2 - \left(\frac{r_v}{a}\right)
%^{n\pi/\beta}\right].
%\end{equation}
and the moment of inertia for rotation about the origin  by $I_V = 
L_V/\Omega$. Using the result for
the vortex position, this can be written as
%CD corrected this equation
\begin{widetext}
\begin{equation}
I_V = \frac{64a^4}{\pi} \frac{1}{[4-\pi^2/\beta^2][(a/r_v)^{\pi/\beta}
-(r_v/a)^{\pi/\beta}]} \nonumber \\ 
\times\sum_{n>0,\; \text{ $n$ odd}} (-1)^{\frac{n+1}{2}} \frac{1}
{n(4-n^2\pi^2 /\beta^2)} \left[\left(\frac{r_v}{a}\right)^2 - 
\left(\frac{r_v}{a}\right)
^{n\pi/\beta}\right].
\label{iveqn}
\end{equation}
\end{widetext}
In the presence of the vortex, the moment of inertia 
about the origin is  $(I_{SF}+I_V)$ where $I_{SF}$ is given by Eq.~(\ref{moi1})
and $I_V$ is given by the equation above. 
The value of $r_v/a$ to be used in this equation is  given by the solution of
Eq.~(\ref{vcond}). Since the vortex position $r_v$ depends on the angular
speed $\Omega$, the value of $I_V$ also depends on $\Omega$. 
%CD deleted several lines below and added new text
%Results for the
%ratios of the moments of inertia of the superfluid in the presence of a vortex
%and that of a rigid body for rotation about the origin and about the COM,
%obtained for $\Omega a^2 = \kappa$, can easily be
%obtained.
%However, they are not shown since
%for typical experimental situations, at $\Omega a^2 = \kappa$ the radius
%of the region where $v$ exceeds $v_c$ would be, as explained above, 
%unphysically small.

%OTV minor rewording below

Although the divergence in the velocity field at small $r$ %OTVfor any nonzero value of $\Omega$ 
is eliminated by the introduction of a 
vortex, the free energy of the state with this vortex is not 
necessarily lower than that of the
vortex-free state with a small region of normal fluid near $r=0$. 
Specifically, in
%CDN Changed wording below
experimental situations (e.g. in experiments on solid $^4{\rm He}$
discussed above)
where the
dimensions of the region of normal fluid are extremely small, 
the free energy cost of creating the
normal region is negligible and the free energy cost 
of creating a vortex is the deciding factor in determining whether a vortex
will be present. We therefore calculate, in the following subsection, 
the free energy of
a state with a single vortex.

%CD Changed title of subsection and text
 %OTV further
\subsection {Free energy of a vortex and critical 
angular velocity for vortex nucleation} 
\label{freeen}

\begin{figure}
\includegraphics[width=3.2in]{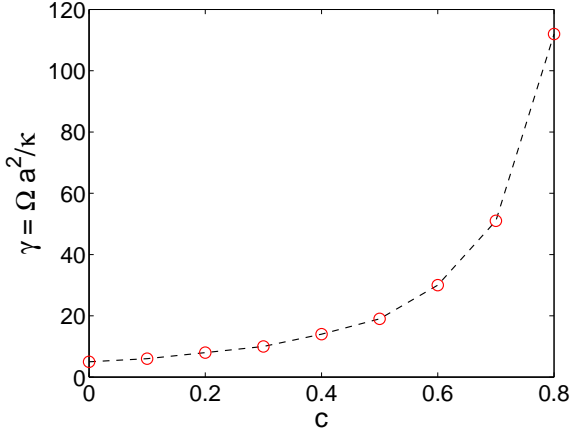}
\caption{The critical angular velocity for
vortex nucleation in a ring ($\beta=2\pi$). Here the critical
value of the parameter $\gamma$ (i.e. $\Omega_1 a^2/\kappa$)
is  plotted as a function of $c$. The circles are numerical
results, connected by straight dashed lines.
The increase at larger $c$ shows  that the nucleation of vortices
is unfavorable in that case.}
\label{fig6}
\end{figure}

%As the angular speed $\Omega$ increases, at a fixed geometry, eventually
%vortices will nucleate. Starting from the vortex-free state, the
%angular speed $\Omega_1$ at which nucleation
%of a first vortex will occur can be determined from free energy
%considerations. 
In the free energy calculation, we consider the general case of a 
ring with $b \ne 0$. 
%OTV
The
angular speed $\Omega_1$ at which nucleation
of a first vortex will occur can be determined from free energy
considerations. 
%OTV As discussed in Ref.~\cite{fetter}, 
The free energy $F$ is given\cite{fetter} in terms of
the energy  $E$ and the angular momentum $L$ as:
\begin{equation}
F=E-L\Omega
\label{free}
\end{equation}
We  will denote here with a subscript 0 the quantities $F$, $E$
and $L$ in the vortex-free state, and with a 1 subscript those in the presence of one 
vortex. As stated in the preceding subsection, 
the stream function in the presence of a vortex is:
\begin{equation}
\Psi_1({\bf r})=\Psi_0({\bf r})+\kappa G({\bf r},{\bf r'}) \equiv  \Psi_0+\Psi_1
\label{psivort}
\end{equation}
where $G({\bf r},{\bf r'})$ is the Green's function given in 
Eq.~(\ref{grfn}) and 
${\bf r'}$ is the vortex position with
coordinates $r',\phi'$. From symmetry considerations $\phi'= \beta/2$
and the equilibrium radial
position of the vortex, $r' = r_v$, is to be determined 
from free energy minimization. The velocity field and the angular momentum in the
presence of a vortex can be readily obtained from the stream 
function of Eq.~(\ref{psivort}).
The angular momentum is given by
\begin{equation}
L_1=L_0+\kappa a^2 C
\label{l1}
\end{equation}
where the dimensionless quantity $C$ has the following expression:
\begin{widetext}
\begin{equation}
C = \frac{8}{\pi} \sum_{n>0,\; \text{ $n$ odd}} (-1)^{\frac{n+1}{2}} \frac{1}{n} \frac{1}{4-x_n^2} \frac{1}{1-c^{2x_n}} \nonumber \\
\times [(r'/a)^2 (1-c^{2x_n})-(r'/a)^{x_n} (1-c^{x_n+2})
-(ca/r')^{x_n}(c^2-c^{x_n}) ],
\label{l2}
\end{equation}
\end{widetext}
with $x_n = n\pi/\beta$.

It is not hard to see explicitly
that $G({\bf r},{\bf r'})$ has, as expected, a logarithmic singularity
at ${\bf r'}$, so that we can write:
\begin{equation}
G({\bf r},{\bf r'})=\frac{1}{2\pi}\ln(|{\bf r-r'}|/\alpha) + g({\bf r},{\bf r'})
\label{sing}
\end{equation}
where $\alpha$ is the radius of the vortex core and $g({\bf r},{\bf r'})$, the nonsingular part
of the Green's function,
satisfies the Laplace equation. 
As shown in Ref.~\cite{fetter} (see also Ref.~\cite{fetter1}), the
energy in the presence of a vortex can be written as
\begin{equation}
E_1=\frac{1}{2}L_1\Omega+\frac{1}{4}\kappa \Omega r'^2-\frac{1}{2}\kappa \Psi_0({\bf r'})-
\frac{1}{2}\kappa^2 g({\bf r',r'}).
\label{e1}
\end{equation}
After some algebra, the nonsingular part of the Green's function appearing in Eq.(\ref{e1}) is 
obtained as
\begin{widetext}
\begin{equation}
g({\bf r',r'})= \frac{1}{2\pi} \ln\left(\frac{\pi \alpha}{2\beta r'}\right) -\frac{1}{\pi} \sum_{n>0,\; \text{ $n$ odd}} 
\frac{1}{n} \frac{1}{1-c^{2x_n}} [2 c^{2x_n} -(r'/a)^{2x_n} - (ca/r')^{2x_n} ],
\label{g00}
\end{equation}
\end{widetext}
where $x_n = n\pi/\beta$. 
Using Eqs.~(\ref{l1}), (\ref{l2}), (\ref{e1}) and (\ref{g00}), 
the free energy in the presence of a vortex at $(r',\beta/2)$ may 
be obtained.
The results depend on the vortex core size, via  the logarithmic
dependence on $a/\alpha$ mentioned above.  One
then minimizes $F_1$ with respect to $r'$ to obtain its
optimal value $r_v$, and 
compares $F_1$ and $F_0$ to find the overall equilibrium state. 
This depends on the value of $\Omega$ and, for sufficiently
small $\Omega$, it  is the vortex-free state, while for
$\Omega > \Omega_1$ 
the one-vortex state first becomes favorable. In practice
these calculations can be done only numerically, but the computations 
are not difficult.
The relevant dimensionless parameter is  the quantity 
$\gamma = \Omega a^2/\kappa$ defined in the preceding section. 
This parameter is the ratio of the characteristic scale,
$\Omega a$, of the velocity field  ${\bf v}^0$ %OTV3 around here
due to the rotation alone, and the scale of the additional velocity
field  ${\bf v}^1$ due to the vortex, which is 
$\kappa/a$. One needs also
to input the value of $\alpha/a$ for which we take the physically
reasonable value of $10^7$ appropriate for liquid $^4{\rm He}$.

Results for $\Omega_1$ computed
for a blocked annular ring  ($\beta=2\pi$) are given in Fig.~\ref{fig6}. There
we plot the critical value of $\gamma$  vs the aspect ratio $c$. 
%The
%critical value has been calculated at the optimum vortex position $r_v$,
%which as explained above depends on $c$ and $\Omega$. 
We see that
at reasonably small or intermediate values of $c$ the critical
value of $\gamma$
is in the range 10-50 corresponding to angular speeds in the general
range of $10^{-1}$/s, which is in the experimentally
relevant region. At large values of $c$ this quantity increases,
reflecting that the system is behaving more like a rigid body, in which
case the formation of vortices is obviously less favorable. A similar
trend was seen for progressively flatter
ellipsoids in Ref.~\onlinecite{fetter}.  
%CDN changed the following sentence
This implies that one need not worry
about the formation of vortices in narrow blocked rings and wedges while
estimating the contribution of these objects to the NCRI of the system.
%the
%conclusions reached in our calculations without vortices about the sharp
%discontinuities that should occur when a reasonably narrow ring
%is obstructed or unobstructed are valid in the experimentally interesting
%range. 

\begin{figure}
\includegraphics[width=3.25in]{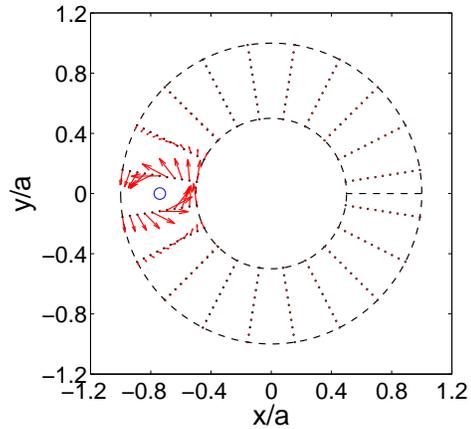}
  %would now be 7
\caption{(Color online) Fields produced by a nucleated vortex in an obstructed
ring with $c=0.5$, at $\Omega=\Omega_1$. Only the fields produced
by the vortex are included. Its position
(marked by a (blue) open dot) is at the
optimal value (see text) $r_v/a=0.74$. The total flow is the
sum of that shown in this figure, weighed by  a factor
of $1/\gamma$, and that
in the top panel of Fig.~\ref{fig1}. Because $\gamma$ is rather large,
the result would be hard to distinguish from that shown in Fig.~\ref{fig1}.  }
\label{fig7}
 %label changed
\end{figure}

In Fig.~\ref{fig7}, we show the texture of
the velocity field ${\bf v}^1$ due
to the nucleated vortex alone  at $c=0.5$ and at  a value of $\gamma$
slightly higher than 
its critical value, which at this value of
$c$
is $\gamma_1 \simeq 20$ (see Fig.~\ref{fig6}). The calculated
optimal position of the vortex at these values of $\gamma$  and $c$ is
$r_v/a=0.74$.  This position  is 
marked by a (blue) circle
in the plot.
The fields in this figure should be
combined with those in the top panel of Fig.~\ref{fig1}. One should recall
that both plots are in arbitrary units, so that
before plotting the combined
field one should divide the fields in Fig.~\ref{fig7} by
$\gamma \simeq 20$ to take into account their overall smaller relative
scale.
%OTVR3B minor edit because of no fig 8 now.
If that were done, however, then the plot would be very hard to
distinguish with the naked eye from  that
in the top panel of Fig.~\ref{fig1}. %For this
%reason, in  order to give the reader a clearer feeling for the
%overall texture of the combined fields, we have plotted them in Fig.~\ref{fig8}
%after multiplying  the ${\bf v}^1$ by a relative weight
%of 1/2, rather than 1/20.  

%\begin{figure}
%\includegraphics[width=4in]{fig8.eps}
 %would now be 8
%\caption {(Color online) Combined velocity fields in
%the presence of a vortex. This is the weighted
%sum of the fields in the top panel in Fig.~\ref{fig1}
%and those in Fig.~\ref{fig7}.  As discussed in the text, the effect of the 
%vortex has been artificially enhanced for visibility
%purposes by using a weight
%factor of 1/2 for the vortex fields, rather than the actual 1/20.}
%\label{fig8}
  %label changed
%\end{figure}

The moment of inertia 
of a ring in the presence of a nucleated vortex may be calculated from
Eqs.~(\ref{l1}) and (\ref{l2}).
 %OTV3
The results deviate from those obtained for the vortex-free state only by a 
correction
of order $1/\gamma$. For $\Omega \ge \Omega_1$ this is therefore
significant only at small values of $c$. At $c \rightarrow 0$ we
find for example  that,  at $\beta=2 \pi$, the moment of inertia of a 
blocked wedge
($c=0$) increases by about 8.3\% as a vortex is nucleated at $\Omega=\Omega_1$,
and the increase in the moment of inertia due to the nucleation of a vortex
becomes less than 1\% for $c \ge 0.33$.

%CD1 New lines below
 %OTV3 minor edits

The optimal value, $r_v$, of the radial coordinate of the
vortex obtained from free-energy minimization is quite different from the value
for which the velocity due to the vortex cancels the mathematical singularity
at $r=0$ found in wedges with $\beta > \pi$. This  implies that the velocity
field would formally diverge at $r=0$ in such systems 
even when a vortex is present
at the position corresponding to the minimum of the free energy. As noted above,
this mathematical singularity does not have any physical consequence in usual
%CDN changed wording below  %OTVN small edit
experiments on $^4{\rm He}$. %We mention this curious fact just to point
%out that 
However, this interplay between the requirements of keeping the velocity below
the Landau critical value and minimizing the free energy may lead to nontrivial
behaviors in other experimentally accessible situations such as 
Bose-Einstein condensates in cold atomic systems. 
%in which the effects of the
%singularity would be observable.

%\begin{figure}
%\includegraphics[width=4.47in]{fig8.eps}
%\caption{Fields produced by a nucleated vortex in an obstructed
%ring with $c=0.5$, at $\Omega=\Omega_1$. Only the fields produced
%by the vortex are included in the plot. The total flow is the
%sum of that shown in this figure, weighed by  a factor
%of $1/\gamma$, and that
%in the top panel of Fig.~\ref{fig1}. Because $\gamma$ is large,
%the result would be hard to distinguish from that shown in Fig.~\ref{fig1}.  }
%\label{fig8}
%\end{figure}

\section{Summary and discussion}
\label{summary}

We have calculated here the velocity fields of a superfluid sample in
a cylindrical wedge, or ring-wedge geometry. We have used two different methods 
to solve the relevant hydrodynamic equations both in the absence of vortices
and when vortices are present. From the resulting velocity fields, 
we have derived formulas for the
moment of inertia, and therefore for the NCRI effect in these geometries. 

Physically, the most important of our results is that the NCRI effect
is most prominent for relatively %OTV2
narrow rings. Our calculations show  %OTV2
that the moment of
inertia of a blocked narrow ring is very close to the rigid-body value unless %OTV2
the width of the ring is a large fraction of its outer radius. %OTV2
Since the moment of inertia
of a superfluid ring for rotation about its center is zero when it is
unblocked (at least for small $\Omega$), one should see a considerable
change in the NCRI when approximately circular superfluid channels in a sample
are obstructed or unobstructed. The fractional change in the moment of inertia 
as a ring is unblocked (defined relative to the moment of inertia of the ring for
rigid-body rotation) is %OTV2
maximum when the rotation axis passes through the
center of the ring. In that case, this ratio approaches unity very quickly as 
the aspect ratio $c$ of the
ring is increased toward one (see Fig.~\ref{fig2}, top panel), 
and this ratio has a value close to 0.44 as 
$c \to 0$. The magnitude of the change in the rotational inertia upon
blocking/unblocking does not depend on the location of the axis of rotation.
For a fixed value of the outer radius $a$, the magnitude of this change is 
maximum when the aspect ratio $c$ is close to 0.52 (see Fig.~\ref{fig2}, bottom
panel). This maximum is very broad. %OTV2
For an annular superfluid wedge, the moment of inertia about an axis
passing through its tip is close
to the rigid-body value if the opening angle $\beta$ is small, and it decreases
as $\beta$ is increased (see Figs.~\ref{fig3} and \ref{fig5}).

%much smaller for a wedge, circular or annular, than for a circle,
%where (at least at small $\Omega$) the moment of inertia is zero. 
%There are several reasons for this: one is that the reduction of the
%moment of inertia of  a wedge with respect to its center (i.e.
%the center of the circles that compose its nonradial boundaries) is always
%smaller than the corresponding reduction with respect to the center
%of mass. But even in the case of a $2\pi$ ring, where the center
%of rotation and the center of mass are the same, there is a very
%considerable difference
%between the moment of inertia found, and that of an unobstructed ring.
%This implies that in a typical experiment one should see a considerable
%change in the NCRI when approximately circular paths centered on the rotation
%axis are obstructed or unobstructed. 

%CD1 changed the wording in the paragraph below
 %OTV3 minor
The results summarized above are for the case where there are no vortices, so that
the velocity field is irrotational. Since one expects
vortices to be nucleated as
the rotational speed in increased,  we have 
used a free-energy  criterion to determine the critical angular
speed for the nucleation of a vortex in the system. 
We find  that in standard ``supersolid'' experiments the %OTVN
relevant range of geometries and speeds includes both 
the parameter region where vortices are
absent and that where nucleated vortices exist. 
For a fixed value of 
$\beta=2\pi$ (ring geometry), the critical angular speed increases
rapidly as the aspect ratio $c$ is increased above about 0.5 (see Fig.~\ref{fig6}).
Also, the increase in the moment of inertia due to the nucleation of a vortex is
rather small (less than 10\%) in all cases.
These observations imply that the results mentioned above for a narrow ring without
vortices remain, for $^4{\rm He}$, %OTVN
valid for relatively large values of the angular velocity.

%CD1 Changed wording below
Mathematically, a number of relevant results have been uncovered and 
emphasized. There are a number of technical difficulties in the calculation
of the velocity fields, leading to non-convergent series and
singularities. However, the singularities are integrable and the series are
Borel summable, so that there is no difficulty in calculating physical 
quantities such as the angular momentum and the kinetic energy. 
%hence
%do not affect the calculation of the moments of inertia, they have been
%largely overlooked. We show that  the integrability of the singularities
%is related to the Borel summability of the asymptotic series occurring
%in the expression for the velocity fields. 
We also point out the occurrence of a mathematical singularity in the velocity 
field in wedges (but not in rings) with $\beta > \pi$ and discuss 
possible effects of this divergence. This singularity turns out to have
%CDN Inserted text below  %OTVN found it
no measurable consequence in experimental studies of $^4{\rm He}$, but may be
relevant in studies of cold atomic systems confined in wedge-shaped traps.

%the effect of vortices occurring because of a mathematical singularity
%at the origin for wedges (not rings) with $\beta> \pi$.

In general, the ideas and methods developed here can be used in other geometries.
%The simpler case of ellipsoids was studied time ago in Ref.~\cite{fetter}.
We believe that the results and techniques presented here can be very
useful in understanding not only NCRI phenomena in ``supersolid'' helium, but
also superflow in confined geometries and in finite systems.
Work in which 
we apply these ideas to study the NCRI effect in realistic
models of grain boundary networks is in progress. %LAST

% tables should appear as floats within the text
%
% Here is an example of the general form of a table:
% Fill in the caption in the braces of the \caption{} command. Put the label
% that you will use with \ref{} command in the braces of the \label{} command.
% Insert the column specifiers (l, r, c, d, etc.) in the empty braces of the
% \begin{tabular}{} command.
% The ruledtabular environment adds doubled rules to table and sets a
% reasonable default table settings.
% Use the table* environment to get a full-width table in two-column
% Add \usepackage{longtable} and the longtable (or longtable*}
% environment for nicely formatted long tables. Or use the the [H]
% placement option to break a long table (with less control than 
% in longtable).
% \begin{table}%[H] add [H] placement to break table across pages
% \caption{\label{}}
% \begin{ruledtabular}
% \begin{tabular}{}
% Lines of table here ending with \\
% \end{tabular}
% \end{ruledtabular}
% \end{table}

% Specify following sections are appendices. Use \appendix* if there
% only one appendix.
%\appendix
%\section{}

% If you have acknowledgments, this puts in the proper section head.
\begin{acknowledgments}
This work was supported  in part by NSF (OISE-0352598) and by
DST (India).
\end{acknowledgments}

% Create the reference section using BibTeX:
%\bibliography{basename of .bib file}
%CDN included several new references

\end{document}